\RequirePackage{etoolbox}\csdef{input@path}{{style/}{graphics/}}
\documentclass[MSNbibl,nameyear,dvips]{arxstspdf}
\usepackage{flushend}
\usepackage{stfloats}

\volume{29}
\issue{4}
\pubyear{2014}
\firstpage{687}
\lastpage{706}
\doi{10.1214/14-STS479} 

\makeatletter
\newproclaim{assumption}{Assumption}
\newcommand{\Prr}{\operatorname{Pr}}
\makeatother

\begin{document}
\begin{frontmatter}

\title{Interference and Sensitivity Analysis}
\runtitle{Interference and Sensitivity Analysis}

\begin{aug}
\author[A]{\fnms{Tyler J.}~\snm{VanderWeele}\corref{}\ead[label=e1]{tvanderw@hsph.harvard.edu}},
\author[A]{\fnms{Eric J.}~\snm{Tchetgen Tchetgen}}
\and
\author[B]{\fnms{M. Elizabeth}~\snm{Halloran}}
\runauthor{T. J. VanderWeele, E. J. Tchetgen Tchetgen and M. E. Halloran}

\affiliation{Harvard School of Public Health, Harvard School of Public Health,
Fred Hutchinson Cancer
Research Center and University of Washington}

\address[A]{Tyler J. VanderWeele is Professor and Eric J. Tchetgen Tchetgen is Associate Professor,
Departments of Epidemiology and Biostatistcs,
Harvard School of Public Health, 677 Huntington Avenue, Boston, Massachusetts 02115, USA \printead{e1}.}
\address[B]{\mbox{M.~Elizabeth} Halloran is Professor,  Fred Hutchinson Cancer
Research Center, 1100 Fairview Ave. N., M2-C200, Seattle, Washington 98109, USA.}
\end{aug}

\begin{abstract}
Causal inference with interference is a rapidly growing
area. The literature has begun to relax the ``no-interference'' assumption
that the treatment received by one individual does not affect the outcomes
of other individuals. In this paper we briefly review the literature on
causal inference in the presence of interference when treatments have been
randomized. We then consider settings in which causal effects  in the
presence of interference are not identified, either because randomization
alone does not suffice for identification or because treatment is not
randomized and there may be unmeasured confounders of the treatment--outcome
relationship. We develop sensitivity analysis techniques for these settings.
We describe several sensitivity analysis techniques for the infectiousness
effect which, in a vaccine trial, captures the effect of the vaccine of one
person on protecting a second person from infection even if the first is
infected. We also develop two sensitivity analysis techniques for causal
effects under interference  in the presence of unmeasured confounding which generalize analogous
techniques when interference is absent. These two techniques for unmeasured
confounding are compared and contrasted.
\end{abstract}

\end{frontmatter}

\section{Introduction}\label{sec1}

Cox [(\citeyear{4}), page 19] wrote that there is no interference between different
units if the observation on one unit is unaffected by the particular
assignment of treatment to the other units. The assumption of no
interference is a key component of Rubin's ``stable unit treatment value
assumption,'' called SUTVA (\cite{27}), that is often required for
potential outcomes to be well-defined. However, in many settings, the
assumption of no interference obviously does not hold. Consider an
individual who, if not vaccinated, would have infected another person, but
who, if vaccinated, would not infect that other person. In this case, the
infection outcome of the second person depends on the treatment of the first
individual, and there is thus interference. Under the assumption of no
interference, the effect of a treatment compares two potential outcomes the
individual would exhibit under treatment and control. With interference, an
individual could have many potential outcomes depending on the treatments
assigned to the other individuals (Rubin, \citeyear{26}, \citeyear{28}).

In some settings interference is a nuisance, while in other settings it
creates effects of scientific, public health or social science interest. An
example of the former includes agricultural experiments where treatments in
neighboring plots can interfere with one another (\cite{15}). Fallow rows
between treatment plots can sometimes eliminate interference between plots,
but more often the interference must be taken into account. In infectious
diseases, interference is inherent in the biology of transmission, it cannot
be eliminated, and it produces intrinsically interesting effects. Social
interaction is a primary source of interference in studies with humans
subjects and often cannot be eliminated.

Progress in causal inference with interference has been made recently in
different contexts, including those in the social sciences, econometrics
and infectious diseases. Several causal effects can be defined in the
presence of interference, and sometimes similar effects have different names
in different contexts. Social scientists have long been interested in the
effects of neighborhoods on the economic, sociological and psychological
well-being of their inhabitants, resulting in the term neighborhood effects
(\cite{30}). The consequences of interference between individuals in this
context are also known as spillover effects. In infectious diseases, these
effects were generally called indirect effects of interventions (\cite{9}).

In Section~\ref{sec1.1} we present informally some examples of studies on causal
inference with interference in different contexts. In Section~\ref{sec2} we present
formal definitions of direct, indirect, total and overall effects as well as
infectiousness effects in the presence of interference. In Section~\ref{sec3} we
develop a number of new sensitivity analysis techniques in settings in which
causal effects are not identified, either because the effect estimand itself
relies on assumptions beyond randomization or because treatment is not
randomized and there may be unmeasured confounding. The sensitivity analysis
techniques help address these issues of identification in these settings.
Section~\ref{sec3} contains the new results of the paper and, as will be seen below,
many of these new results in the context of interference build on foundational work on sensitivity analysis by
\citet{24} outside the context of interference. Section~\ref{sec4}
offers some concluding remarks on directions for future research on
interference. A reader who is primarily interested in the technical
development can skip Section~\ref{sec1.1} and move on directly to Section~\ref{sec2}.

\subsection{Motivating Examples}\label{sec1.1}

\subsubsection{Interference and housing mobility}\label{sec1.1.1}

Sobel\break  (\citeyear{30}) considered interference in the Moving to Opportunity (MTO)
demonstration sponsored by the U.S. Department of Housing and Urban
Development. In this housing mobility experiment in poor neighborhoods in
five cities, eligible ghetto residents were randomly assigned to receive one
of two forms of relocation assistance or no assistance (control). Sobel
argued that the no interference assumption is not plausible for the MTO
demonstration because many of the participants likely knew other
participants at each of the five sites. Thus, the participants could have
influenced each other through social interaction. For example, a family that
decided to move to a new neighborhood could give rise to worse outcomes for
a family that stayed in the original neighborhood because of the decline in
social support for the family that stayed.

\citet{30} defined causal estimands and estimators for indirect/spillover
effects for the MTO randomized trial of housing vouchers, taking compliance
into account. He assumed that interference could occur within the sites, but
not across sites, which he called partial interference. He made a key
contribution in proposing causal estimands for assessing effects in the
presence of interference by averaging causal effects over all possible
treatment assignments for a particular allocation strategy compared to a
benchmark strategy wherein no units received the treatment assignment.
Although his language is different, he essentially defined causal estimands
analogous to the direct, indirect, total and overall effects defined in the
next section.

He then compared his causal estimands to what is usually estimated in
studies of housing mobility not taking interference into account. He showed
that what is usually estimated actually gives the difference between (i) the
average effect of the voucher on those who received them and (ii) the
average effect on those not receiving vouchers of having people leave the
neighborhood. Both effects could be negative (detrimental) with the
difference positive, thus making it important to take potential interference
into account.

\begin{table*}[b]
\tablewidth=\textwidth
\tablewidth=12cm
\tabcolsep=0pt
\caption{Illustrative example of a two-stage randomized placebo-controlled
cholera vaccine trial based on data from Ali et~al. (\citeyear{1}). Group
assignment corresponds to 50\% or 30\% vaccine coverage (from Hudgens and Halloran (\citeyear{13}))}\label{tab:cholera1}
\begin{tabular*}{12cm}{@{\extracolsep{4in minus 4in}}lccccc@{}}
\hline
& & \multicolumn{2}{c}{\textbf{Vaccine recipients}} &   \multicolumn{2}{c@{}}{\textbf{Placebo recipients}} \\[-5pt]
& \textbf{Group  assignment}& \multicolumn{2}{c}{\hrulefill} &   \multicolumn{2}{c@{}}{\hrulefill}\\
\textbf{Group} & \textbf{(\% vaccinated)} & \textbf{Total}             & \textbf{Cases} & \textbf{Total}         & \multicolumn{1}{c@{}}{\textbf{Cases}}          \\\hline
1     & 50              & 12\mbox{,}541     & 16    & 12\mbox{,}541 & \hphantom{1}18 \\
2     & 50              & 11\mbox{,}513     & 26    & 11\mbox{,}513 & \hphantom{1}54 \\
3     & 30              & 10\mbox{,}772     & 17    & 25\mbox{,}134 & 119            \\
4     & 30              & \hphantom{1,}8883 & 22    & 20\mbox{,}727 & 122            \\
5     & 30              & \hphantom{1,}5627 & 15    & 13\mbox{,}130 & \hphantom{1}92       \\
\hline
\end{tabular*}
\end{table*}

\subsubsection{Interference in vaccination programs}\label{sec1.1.2}

Motivated by an interest in the effects of vaccination and vaccination
programs, \citet{31} and Halloran and
Struchiner (\citeyear{9}, \citeyear{10}) conceptually defined direct, indirect, total and
overall effects in the presence of interference. The direct effect of a
treatment on an individual was defined as the difference between the
potential outcome for that individual given treatment compared to the
potential outcome for that individual without treatment if the treatment
assignment in the others in the population was held fixed. In contrast to
direct effects, an indirect effect describes the effect on an individual of
the treatment received by others in the group when that individual's
treatment was held fixed. In particular, the indirect effect of a treatment
on an individual was defined as the difference between the potential
outcomes for that individual without treatment when the group (i) receives
an intervention program and (ii) receives a benchmark program of no
intervention. Total effects describe the combination of direct and indirect
effects of a particular treatment assignment on an individual. The total
effect of a treatment on an individual is the difference between the
potential outcomes for that individual (i) with treatment when the group
receives an intervention program and (ii) without treatment when the group
receives no intervention. Overall effects describe the average effect of an
intervention relative to no intervention.

\citet{10} proposed individual-level causal estimands of
direct, indirect, total and overall in the presence of interference by
letting the potential outcomes for any individual depend on the vector of
treatment assignments to other individuals in the group (Rubin, \citeyear{26}, \citeyear{28}).
However, they did not propose population level causal estimands.

A number of studies have been conducted to estimate indirect, total or
overall effects of vaccination programs outside of the causal inference
framework. In the United Kingdom, the indirect effect of a new program of
meningococcal C vaccination was estimated by comparing the attack rates in
unvaccinated children and adolescents before and after introduction of the
program (\cite{23}). The United Kingdom
introduced routine meningococcal serogroup C vaccination for infants in
November 1999. The vaccine was also offered to all children and adolescents
aged $<$18 years in a phased catch-up program. Adolescents were vaccinated
first and the program was completed by the end of 2000. About 75\% of the
children and adolescents were vaccinated. The attack rate in \textit{%
unvaccinated} infants through adolescents per 100,000 unvaccinated
population in July 1998--June 1999 was 4.08 (95\% CI: 3.7, 4.5) and in July
2001--June 2002 was 1.36 (95\% CI: 0.86, 1.85). Vaccinating about 75\% of the
children and adolescents thus seemed to produce an indirect effect, with a
relative reduction in the number of confirmed meningococcal C cases in the
unvaccinated children and adolescents, of 67\% (95\% CI: 52, 77).


To obtain group- and population-level causal estimands for direct, indirect,
total and overall causal effects of treatment, \citet{13}
proposed a two-stage randomization scheme, the first stage at the group
level, the second at the individual level within groups based on Sobel's
approach of averaging over all possible treatment assignments.
As did \citet{30}, they assumed interference can occur within groups but
not across groups. The causal estimands defined by \citet{13} are applicable to other situations with interference in fixed groups
of individuals where treatment can be assigned to individuals within groups.
A brief formal development is given in Section~\ref{sec2}.

As an example, \citet{13} presented a hypothetical
two-stage randomized placebo-controlled trial of cholera vaccines (Table~\ref{tab:cholera1}).
Suppose in the first stage five geographically separate groups were
randomized 
so two were assigned to vaccinate 50\% and three were assigned to vaccinate
30\% of individuals, then individuals were randomly assigned to be
vaccinated or not. Causal effect estimates (and estimated variance) are given in
the change in number of cases per 1000 individuals per year. The estimated
indirect effect of vaccinating 50\% versus 30\% in the unvaccinated
individuals is 2.81 (3.079). This suggests that vaccinating 50\% of the
population would result in 2.8 fewer cases per 1000 unvaccinated people per
year compared with vaccinating only 30\%. Similarly, the estimated total
effect is 4.11 (0.672). This suggests that vaccinating 50\% of the
population would result in about 4.1 fewer cases per 1000 vaccinated people per
year compared with unvaccinated persons vaccinating only 30\%. The estimated overall effect is
2.37 (1.430). The estimated overall effect is a summary comparison of the
two strategies, suggesting that, on average, 50\% vaccine coverage results
in 2.4 fewer cases of cholera per 1000 individuals per year compared to 30\%
vaccine coverage. A public health professional could use these estimates in
evaluating the cost-benefit of vaccinating more people and preventing more
cases versus vaccinating fewer people. The direct effect under 30\% coverage
is 3.64 (0.178), nearly three times greater than the direct effect under
50\% coverage, which is 1.30 (0.856). The difference shows that even the
direct effects can depend on the level of coverage due to interference
between individuals. Note that the direct effect under 50\% coverage of 1.30
and the indirect effect of 2.81 sum to the total effect of 4.11.

\subsubsection{Interference in the context of kindergarten retention}\label{sec1.1.3}

Hong and Raudenbush (\citeyear{11}) considered interference in the context of the
effect on reading scores of children of being retained in kindergarten
versus being promoted to the first grade. Interference was assumed possible
through the dependence of the potential outcomes of reading test scores of
one child on whether other children were retained or not. Hong and
Raudenbush were principally interested in the effect of a child's being
retained and how this varied with being in schools with low retention and
versus those with high retention. They used a sample of data from 1080
schools with 471 kindergarten retainees and 10,255 promoted students. In
their application, students are clustered in schools. Individual treatment
assignment was whether a student is retained. They used a school-level
scalar function based on the proportion of the students that were retained
to determine whether a school was a ``high-retention'' or ``low-retention''
school. The study was observational at two levels: schools were not
randomized to have high or low retention, and students were not randomized
to be retained. However, they framed their analysis within a two-stage
randomization procedure similar to that described in \citet{13} in which both stages would have been randomized. They also assumed
interference was possible within schools but not across schools.

Using a propensity score-based approach, accounting for interference, and
assuming that assignment at both the school and the individual level was
ignorable given a number of observed individual-level, school-level, and
school-aggregated-individual-level characteristics, \citet{11} obtained estimates of the effect on reading scores of retention in
high-retention and low-retention schools. Specifically, in low-retention
schools, they estimated the effect on reading scores of a student being
retained versus being promoted, was $-$8.18 (95\% CI: $-$10.02, $-$6.34), and in
high-retention schools the effect estimate was $-$8.86 (95\% CI: $-$11.56, $-$6.16).
A standard deviation in reading test scores in this sample is 13.48
points. We will return to this example below to demonstrate sensitivity
analysis in the context of interference.

\subsubsection{Interference between two sides of the face}\label{sec1.1.4}

Rosenbaum (\citeyear{25}) took a different approach to causal inference with
interference when analyzing randomized experiments than those in previous
sections. He pointed out that if Fisher's null hypothesis of no effect for
any individual in the population is true, then there is no effect and
consequently no interference. Thus, Fisher's permutation test of no effect
will have the correct level, even if, under the alternative hypothesis,
there would be interference. He presented several examples, including data
from a randomized, double-blind experiment in which 15 people received
different preparations of botulinum A exotoxin on each side of their face to
treat wrinkles to test which was less painful. 

Rosenbaum presented exact nonparametric methods for inverting randomization
tests to obtain confidence intervals for assessing treatment effect assuming
nothing about the structure of the interference between units. He assumed
that there were a number of blocks (groups) with a number of individuals
within each group, some of which, but not all, were randomized to a
treatment, the others to control. He developed a general notation that
allowed interference across blocks and did not assume a two-stage
randomization. 
\citet{25} differentiated two null hypotheses. The first null
hypothesis is that treatment has no primary effect, that is, the response of
each unit does not vary under different randomization assignments in the
collection of the possible assignment matrices with fixed number randomized
in each block. Analogous to the benchmark allocation of \citet{30} and the
two-stage randomized trials described above where possibly some communities
receive only the control intervention, \citet{25} invoked uniformity
trials in which individuals within treatment groups would be randomly
assigned to treatment and control, but everyone in control groups would
receive just control. The second null hypothesis is that treatment has no
effect, that is, under different randomization assignments in the collection
of the possible assignment matrices with fixed number randomized in each
block, each individual's response equals his response in a uniformity trial.
If there is no effect, then no benefit is gained from receiving the
treatment. If there is no primary effect, there is no advantage to being one
of the treated individuals, but the benefits could be shared by all of the
individuals.


Rosenbaum gave conditions using distribution-free tests in which without
performing the uniformity trials, he was able to get confidence statements
about the magnitude of the effect and/or primary effect, though not able to
distinguish between them. In the botox example, the 15 people are the
blocks, the two sides of the face the individuals. All 15 people reported
less pain from the treatment containing alcohol. Using his method, the
hypotheses of no effect and no primary effect were rejected with a one-sided
significance level 0.000031.

\citet{18} extended this approach in the context of a cognitive
neuroscience experiment in which the brains of a moderate number of subjects
are studied using functional magnetic resonance imaging while challenged
with a rapid fire sequence of randomized stimuli. Interference was assumed
to occur between units of time in the same individual.

\subsubsection{Interference and infectiousness effects}\label{sec1.1.5}

In vaccine contexts, a vaccinated person who becomes infected might have a
lower probability of transmitting to a susceptible person during a contact
than an unvaccinated person who becomes infected. This is called the effect
of the vaccine on infectiousness. In a study in Niakhar, Senegal, for
example, \citet{22} estimated the relative reduction in
infectiousness to household contacts of a vaccinated case of pertussis
compared to an unvaccinated case to be 67\% (95\% CI:  29, 86). Estimating
reduction in infectiousness can be of considerable public health interest,
particularly with vaccines that do not protect well against infection.

Developing general methods for causal inference for infectiousness effects
poses complicated challenges. Even if the vaccine is randomized, the
infectiousness effect is measured only in people who become infected, a
post-randomization variable, so the estimate would in general be subject to
selection bias. \citet{41} and Halloran and
Hudgens (\citeyear{7}, \citeyear{8}) proposed causal quantities corresponding to the
infectiousness effect in the simple situation of households of size two. The
general approach combines causal inference with interference with principal
stratification (\cite{5}). The latter accounts for the fact
that the comparison in the groups who become infected may be subject to
selection bias. The causal infectiousness effect is not identifiable without
further assumptions. In Section~\ref{sec2.6} we present the bounds that were
developed previously. In Section~\ref{sec3.2} we present new results for sensitivity
analyses for causal infectiousness effects.


\subsubsection{Other approaches}\label{sec1.1.6}

Manski (\citeyear{19}) studied identification of potential outcome distributions when
treatment response may have social interactions. He called the no
interference assumption the \textit{individualistic treatment response} to
differentiate it from other forms of treatment response that depend on
social interaction. \citet{42} discussed the relation
between causal interactions and interference and how under randomization it
is possible to test for specific forms of interference. They show that the
theory for causal interactions provides a conceptual apparatus for assessing
interference as well.

\section{Formalization}\label{sec2}

In this section we present previously developed formalizations of the
direct, indirect, total and overall effects as well as the infectiousness
effects as background for the development of the new sensitivity analyses
under interference in Section~\ref{sec3}.

\subsection{Notation}\label{sec2.1}

Suppose there are $N\geq 1$ groups of individuals or blocks of units. For $%
i=1,\ldots ,N$, let $n_{i}$ denote the number of individuals in group $i$
and let $\mathbf{Z}_{i}=(Z_{i1},\ldots ,Z_{in_{i}})$ denote the treatments
those $n_{i}$ individuals receive. Assume $Z_{ij}$ is a dichotomous random
variable having values 0 or 1 such that $\mathbf{Z}_{i}$ can take on $%
2^{n_{i}}$ possible values. Let $\mathbf{Z}_{i(j)}$ denote the $n_{i}-1$
subvector of $\mathbf{Z}_{i}$ with the $j$th entry deleted. The vector $%
\mathbf{Z}_{i}$ is referred to as an intervention or treatment \textit{%
program}, to distinguish it from the individual treatment $Z_{ij}$. Let $%
\mathbf{z}_{i}$ and $z_{ij}$ denote possible values of $\mathbf{Z}_{i}$ and $%
Z_{ij}$. Define $R^{j}$ to be the set of vectors of possible treatment
programs of length $j$, for $j=1,2,\ldots ,n_{i}$. For example, $R^{2} =
\{(0,0), (0,1), (1,0), (1,1)\}$.

Denote the potential outcome of individual $j$ in group $i$ under treatment $%
\mathbf{z}_{i}$ as $Y_{ij}(\mathbf{z}_{i})$. Denote $\mathbf{Y}_{i}(\mathbf{z%
}_{i})$ as the vector of such outcomes under treatment $\mathbf{z}_{i}$ for
group $i$. The notation $Y_{ij}(\mathbf{z}_{i})$ allows for the possibility
that the potential outcome for the individual $j$ may depend on another
individual's treatment assignment in group $i$, that is, it allows for
interference between individuals within a group. The $Y_{ij}(\mathbf{z}_{i})$
potential responses can be assumed fixed, since they do not depend on the
realized random assignment of treatments $\mathbf{Z}_{i}$, whereas the
observed responses $Y_{ij}(\mathbf{Z}_{i})$ do depend on $\mathbf{Z}_{i}$
and thus are random variables. We also consider potential outcomes $\mathbf{Y%
}_{i}(\mathbf{z}_{i})$ that are independent and identically distributed
across blocks. Partial interference is assumed to hold, that is, the outcome
of one individual can depend on treatment of other individuals in the same
block, but not those in different blocks. The form of the interference
within groups is assumed unknown and can be of arbitrary form.

\subsection{Treatment Assignment Mechanisms}\label{sec2.2}

Following \citet{13}, consider a two-stage randomization
scheme, the first stage at the group level, the second at the individual
level within groups. Let $\psi $ and $\phi $ denote parameterizations that
govern the distribution of $\mathbf{Z}_{i}$ for $i=1,\ldots ,N$.
Corresponding to the first stage of randomization, let $\mathbf{S}\equiv
(S_{1},\ldots ,S_{N})$ denote the group assignments with $S_{i}=1$ if the
group is assigned to $\psi $ and 0 if assigned to $\phi$. Let $\nu $ denote
the parameterization that governs the distribution of $\mathbf{S}$ and let $%
C\equiv \sum_{i}S_{i}$ denote the number of groups assigned $\psi $.
Following \citet{30}, \citet{13} focused on a mixed group
and mixed individual assignment strategy, whereby a fixed number of groups
were allocated to $\psi $, and within each group, a fixed number of
individuals were allocated to treatment versus control. \citet{40} and \citet{32}
considered what we call a simple randomization scheme whereby treatment is
randomly assigned to different individuals within group $i$ according to a
Bernoulli probability mass function.
The causal estimands defined below have the same form under either
randomization scheme, though the different randomization schemes result in
subtle differences of interpretation.

\subsection{Average Potential Outcomes}\label{sec2.3}

Causal estimands are typically defined in terms of averages of potential
outcomes which are identifiable from observable random variables. Following
this approach, the potential outcomes for individual $j$ in group $i$ under $%
z_{ij}=z$ can be written
\begin{eqnarray}
\label{equ1} 
Y_{ij}(\mathbf{z}_{i(j)},z_{ij}=z),
\end{eqnarray}%
for $z=0,1$. Because (\ref{equ1}) depends on $\mathbf{z}_{i(j)}$, following \citet{30}, \citet{13} defined the \textit{individual average
potential outcome} for individual $j$ in group $i$ under $z_{ij}=z$ by
\begin{eqnarray*}
&&\overline{Y}_{ij}(z;\psi )\equiv \sum_{\boldsymbol{\omega }\in
\{0,1\}^{n_{i}-1}}Y_{ij}(
\mathbf{z}_{i(j)}=\boldsymbol{\omega },z_{ij}=z)
\\
&&\hspace*{98pt} {}\cdot {\Prr}_{\psi }(\mathbf{Z}_{i(j)}=
\boldsymbol{\omega }|Z_{ij}=z).
\end{eqnarray*}
In other words, the individual average potential outcome is the conditional
expectation of $Y_{ij}(\mathbf{Z}_{i})$ given $Z_{ij}=1$ under assignment
strategy $\psi $. In contrast, under the simple allocation strategy of
\citet{40}, the potential outcomes are
averaged over the unconditional distribution of $\mathbf{Z}_{i(j)}$.
Averaging over individuals, define the \textit{group average potential
outcome} under treatment assignment $z$ as $\overline{Y}_{i}(z;\psi )\equiv
\sum_{j=1}^{n_{i}}\overline{Y}_{ij}(z;\psi )/n_{i}$. Finally, averaging over
groups, define the \textit{population average potential outcome} under
treatment assignment $z$ as $\overline{Y}(z;\psi )\equiv \sum_{i=1}^{N}%
\overline{Y}_{i}(z;\psi )/N$. The causal estimands in the next section are
defined in terms of the individual, group and population average potential
outcomes. The individual estimands were defined in \citet{10}, the individual average, group average and population average
estimands in \citet{13}.

\subsection{Direct, Indirect, Total and Overall Causal Effects}\label{sec2.4}

The \textit{individual direct causal effects} of treatment 0 compared to
treatment 1 for the individual $j$ in group $i$ were defined by
\begin{eqnarray}
\label{equ2} CE_{ij}^{D}(\mathbf{z}_{i(j)})&\equiv&
Y_{ij}(\mathbf{z}%
_{i(j)},Z_{ij}=1)
\nonumber
\\[-8pt]
\\[-8pt]
&&{}-Y_{ij}(\mathbf{z}_{i(j)},Z_{ij}=0).
\nonumber
\end{eqnarray}%
The \textit{individual average direct causal effect} for the $j$th
individual in the $i$th group was defined by
\begin{equation}
\label{equ3} \overline{CE}_{ij}^{D}(\psi )\equiv
\overline{Y}_{ij}(1;\psi )-\overline{Y}%
_{ij}(0;\psi
), 
\end{equation}%
that is, the difference in individual average potential outcomes when $z_{ij}=1$
and when $z_{ij}=0$ under $\psi $. 
The \textit{group average direct causal effect} as defined by $\overline{CE}%
_{i}^{D}(\psi )\equiv \overline{Y}_{i}(1;\psi )-\overline{Y}_{i}(0;\psi
)=\sum_{j=1}^{n_{i}}\overline{CE}_{ij}^{D}(\psi )/\allowbreak n_{i}$, and the \textit{%
population average direct causal effect} by $\overline{CE}^{D}(\psi )\equiv
\overline{Y}(1;\psi )-\overline{Y}(0;\psi )=\sum_{i=1}^{N}\overline{CE}%
_{i}^{D}(\psi )/N$.

%


The \textit{individual indirect causal effects} of treatment program $%
\mathbf{z}$ compared with $\mathbf{z}^{\prime }$ on individual $j$ in group $%
i$ were defined by
\begin{eqnarray}
\label{equ4} CE_{ij}^{I}\bigl(\mathbf{z}_{i(j)},
\mathbf{z}_{i(j)}^{\prime }\bigr)&\equiv& Y_{i}(%
\mathbf{z}_{i(j)},z_{ij}=0)
\nonumber
\\[-8pt]
\\[-8pt]
&&{}-Y_{i}\bigl(\mathbf{z}_{i(j)}^{\prime },z_{ij}^{\prime
}=0
\bigr),
\nonumber
\end{eqnarray}%
where $\mathbf{z}^{\prime }$ is another $n_{i}$ dimensional vector of
treatment random variables. (Note $\mathbf{z}^{\prime }$ does not denote the
transpose of $\mathbf{z}$.) Similar to direct effects,
the \textit{individual average indirect causal effect} were defined by $%
\overline{CE}_{ij}^{I}(\phi ,\psi )\equiv \overline{Y}_{ij}(0;\phi )-%
\overline{Y}_{ij}(0;\psi )$. Clearly, if $\psi =\phi $, then $\overline{CE}%
_{ij}^{I}(\phi ,\psi )=0$, that is, there will be no individual average
indirect causal effects. Finally, 
the \textit{group average indirect causal effect} was defined as $\overline{%
CE}_{i}^{I}(\phi ,\psi )\equiv\vspace*{2pt} \overline{Y}_{i}(0;\phi )-\overline{Y}%
_{i}(0;\psi )=\sum_{j=1}^{n_{i}}\overline{CE}_{ij}^{I}(\phi ,\allowbreak\psi )/ n_{i}$
and the \textit{population average indirect causal  effect} as $\overline{CE}%
^{I}(\phi ,\psi )\equiv \overline{Y}(0;\phi )-\overline{Y}(0;\psi
)= \sum_{i=1}^{N}\overline{CE}_{i}^{I}(\phi ,\psi )/N$.

%


The \textit{individual total causal effects} for individual $j$ in group $i$
were defined as
\begin{eqnarray}
\label{equ5} CE_{ij}^{T}\bigl(\mathbf{z}_{i(j)},
\mathbf{z}_{i(j)}^{\prime }\bigr)&\equiv& Y_{ij}(%
\mathbf{z}_{i(j)},z_{ij}=1)
\nonumber
\\[-8pt]
\\[-8pt]
&&{}-Y_{ij}\bigl(\mathbf{z}_{i(j)}^{\prime
},z_{ij}^{\prime }=0
\bigr).
\nonumber
\end{eqnarray}%
The \textit{individual average total causal effect} was defined by $%
\overline{CE}_{ij}^{T}(\phi ,\psi )\equiv \overline{Y}_{ij}(1;\phi )-%
\overline{Y}_{ij}(0;\psi )$, the \textit{group average total causal effect}
was defined by $\overline{CE}_{i}^{T}(\phi ,\psi )\equiv \overline{Y}%
_{i}(1;\phi )-\overline{Y}_{i}(0;\psi )=\sum_{j=1}^{n_{i}}\overline{CE}%
_{ij}^{T}(\phi ,\psi )/\break n_{i}$, and the \textit{population average total
causal effect} was defined by $\overline{CE}^{T}(\phi ,\psi )\equiv
\overline{Y}(1;\phi )-\overline{Y}(0;\psi )= \sum_{i=1}^{N}\overline{CE}%
_{i}^{T}(\phi , \psi )/N$. It follows by simple addition and subtraction that
a total effect is the sum of the direct and indirect effects at the
individual, individual average, group average and population average levels.
For example, $\overline{CE}^{T}(\phi ,\psi )=\overline{Y}(1;\phi )-\overline{%
Y}(0;\psi )=\overline{Y}(1;\phi )-\overline{Y}(0;\phi )+\overline{Y}(0;\phi
)-\overline{Y}(0;\psi )=\overline{CE}_{i}^{D}(\phi )+\overline{CE}^{I}(\phi
,\psi )$.

%


The overall causal effect was defined to be the average effect of an
intervention program relative to no intervention.
The \textit{individual overall causal effect} of treatment $\mathbf{z}_{i}$
compared to treatment $\mathbf{z}_{i}^{\prime }$ for individual $j$ in group
$i$ was defined by $CE_{ij}^{O}(\mathbf{z}_{i},\mathbf{z}_{i}^{\prime
})\equiv Y_{ij}(\mathbf{z}_{i})-Y_{ij}(\mathbf{z}_{i}^{\prime })$.
Similarly, for the comparison of $\phi $ to $\psi $, the \textit{individual
average overall causal effect} was defined by $\overline{CE}_{ij}^{O}(\phi
,\psi )\equiv \overline{Y}_{ij}(\phi )-\overline{Y}_{ij}(\psi )$, the
\textit{group overall causal effect} by $\overline{CE}_{i}^{O}(\phi ,\psi
)\equiv \overline{Y}_{i}(\phi )-\overline{Y}_{i}(\psi )$, and the \textit{%
population overall causal effect} by $\overline{CE}^{O}(\phi ,\psi )\equiv
\overline{Y}(\phi )-\overline{Y}(\psi )$ where $\overline{Y}(\psi )\equiv
\sum_{i=1}^{N}\overline{Y}_{i}(\psi )/N$ and~$\overline{Y}_{i}(\psi )\equiv
\sum_{j=1}^{n_{i}}\overline{Y}_{ij}(\psi )/n_{i}$ and $\overline{Y}_{ij}(\psi )\equiv \sum_{\boldsymbol{\omega }\in
\{0,1\}^{n_{i}}}Y_{ij}(\mathbf{z}_{i}=\boldsymbol{\omega }){\Pr }%
_{\psi }(\mathbf{Z}_{i}=\boldsymbol{\omega })$.  \citet{40} showed that the overall effect decomposes into the sum of an
indirect effect and a contrast of two direct effects on the individual
average, group average and population average levels. For example,
$\overline{CE}^{O}(\phi ,\psi )=\overline{CE}^{I}(\phi ,\psi )+\{\overline{CE}%
^{D}(\phi ){\Pr }_{\phi }(Z_{ij}=1\mathbf{)}-\overline{CE}^{D}(\psi ){\Pr }%
_{\psi }(Z_{ij}=1\mathbf{)}\}$.

The quantities defined above under interference have two important
distinctions from those used in causal inference without interference. First,
they quantify causal effects only for participants in the randomized study.
Second, they depend on the randomization probabilities [through $\Pr_{\psi
}(Z_{i(j)}=\omega |Z_{ij}=z)$]. Although the causal estimands here depend on
the assignment mechanism (e.g., comparing two different proportions
vaccinated), we could alternatively compare allocation strategies of always
vaccinate versus never vaccinate to recover traditional causal estimands
that do not depend on the assignment mechanism.

The estimands defined above simplify under the assumption of no interference
between individuals within a group since the potential outcomes of the $%
j${th} individual in group $i$ can be written as $Y_{ij}(1)$ and $Y_{ij}(0)$%
. In turn, the individual direct causal effect is no longer dependent on the
treatment assignment vector $\mathbf{z}_{i(j)}$ and simply equals $%
Y_{ij}(1)-Y_{ij}(0)$. The corresponding group average direct causal effect
becomes $\sum_{j=1}^{n_{i}}\{Y_{ij}(1)-Y_{ij}(0)\}/{n_{i}}$, that is, the usual
average causal effect estimand. By (\ref{equ4}), the individual indirect causal
effect equals zero for all individuals assuming no interference. That is,
assuming no interference implies the treatment has no indirect effects.
Similarly, by (\ref{equ2}) the individual total causal effect equals the individual
direct causal effect. Likewise, at the group average level, under the no
interference assumption the indirect causal effect is zero and the direct
causal effect equals the total causal effect.

\subsection{Inference and Challenges}\label{sec2.5}

Assuming the two-stage randomization and mixed allocation strategy,
\citet{13} proposed unbiased estimators for the various population average
effects. They provided variance estimates under the assumption of stratified
inference, that is, if it matters only how many people are allocated to
treatment, not exactly which ones. \citet{32}
provided conservative variance estimators (i.e., guaranteed to be no smaller
than the true variance in expectation), under more general assumptions and
provided finite sample confidence intervals for the various effects without
the assumption of stratified interference. \citet{16} further
developed large sample randomization inference for the direct, indirect,
total and overall causal effects in the presence of interference when
either the number of groups or the number of individuals within groups grows
large, but not necessarily both.



\subsection{Interference and Infectiousness Effects}\label{sec2.6}

To develop causal estimands for the infectiousness effects presented in
Section~\ref{sec1.1.5}, we follow the development of \citet{41} and \citet{7}. Consider a setting with $%
N $ households (groups) indexed by $i=1,\ldots,N$. Each household consists of
two persons indexed by $j=1,2$. We let $Z_{ij}$ denote the vaccine status
for individual $j$ in household $i$, where $Z_{ij}=1$ if the individual
received vaccine and $Z_{ij}=0$ if the individual did not. For each
household, $\mathbf{Z}_{i}=(Z_{i1},Z_{i2})$ denotes the vaccine status of
the two individuals in the household.  We let $Y_{ij}$ denote the infection
status of individual $j$ in household $i$ after some suitable follow-up in
the study.  We let $Y_{ij}(z_{i1},z_{i2})$ denote the potential outcome for
individual $j$ in household $i$ if the two individuals in that household $i$
had vaccine status of $(z_{i1},z_{i2})$; we treat the potential outcome
vector $\mathbf{Y}_{i}(z_{i1},z_{i2})$ as a random variable that is
independent and identically distributed across households.

We assume partial interference, that is, the exposure status of persons in
one household in the study do not affect the outcomes of individuals in
other study households. The assumption that clusters constitute isolated
pairs would be reasonable in a vaccine trial conducted with a relatively
small number of households in a very large city so that it is unlikely that
the various households in the study would interact with one another.  We
will assume that the two individuals in each household are distinguishable
(e.g., a husband and wife pair) and we will consider a simple randomized
experiment in which only one of the two individuals (e.g., the wife) is
predetermined to be randomized to receive a vaccine or control and the
second person (e.g., the husband) is predetermined to be always unvaccinated.
We let $j=1$ denote the individual who may or may not be vaccinated (e.g.,
the wife) and $j=2$ the individual who is always unvaccinated (e.g., the
husband). In other settings in which the individual (husband or wife) who is
subject to vaccination is itself randomized (i.e., two-stage randomization),
the analysis below could be done separately in those households in which the
wife was selected for vaccine randomization versus those in which the
husband was selected.

The crude (or net) estimator for the infectiousness effect on the risk
difference scale was defined as%
\begin{eqnarray}
\label{equ6} &&E[Y_{i2}|Z_{i1}=1,Y_{i1}=1]
\nonumber
\\[-8pt]
\\[-8pt]
&&\quad{}-E[Y_{i2}|Z_{i1}=0,Y_{i1}=1],
\nonumber
\end{eqnarray}%
where the expectation is taken over all households. This is a comparison of
the infection rates for individual $2$ in the subgroup in which individual $%
1 $ was vaccinated and infected versus in the subgroup in which individual $1
$ was unvaccinated and infected. 
Even though the vaccine status for individual $1$ is randomized,
conditioning on a variable that occurs after treatment, for example, the infection
status of individual $1$, in effect breaks randomization. The net estimator
for the infectiousness effect could be subject to selection bias. We are
computing infection rates for individual $2$ for subpopulations that may be
quite different with respect to individual $1$.

Consider a second contrast proposed by VanderWeele and Tchetgen Tchetgen(\citeyear{41}) and \citet{7}:
\begin{eqnarray}
\label{equ7} &&E\bigl[Y_{i2}(1,0)-Y_{i2}(0,0)|
\nonumber
\\[-8pt]
\\[-8pt]
&&\quad Y_{i1}(1,0)=Y_{i1}(0,0)=1\bigr].
\nonumber
\end{eqnarray}%
This contrast compares the infection status for individual $2$ if individual
$1$ was vaccinated, $Y_{i2}(1,0)$, versus unvaccinated, $Y_{i2}(0,0)$, but
only among the subset of households for whom individual $1$ would have been
infected irrespective of whether individual $1$ was vaccinated, that is, $%
Y_{i1}(1,0)=Y_{i1}(0,0)=1$. Such a subgroup is sometimes called a principal
stratum (\cite{5}). The contrast in (\ref{equ7}) is not subject to
selection bias, so it can be considered a formal causal contrast for the
infectiousness effect.

Unfortunately, we do not know which households fall into the subpopulation
in which individual $1$ would have been infected irrespective of whether
individual $1$ was vaccinated. The contrast (\ref{equ7}) is, in general,
unidentified, even when treatment is randomized, though the observable data
do provide some information about (\ref{equ7}). Bounds and sensitivity analysis are
further facilitated by other assumptions.

\begin{assumption}\label{ass1}
For all $i$, $Y_{i1}(1,0)\leq Y_{i1}(0,0)$.
\end{assumption}

Assumption~\ref{ass1}, usually called a monotonicity assumption, states there is no
one who would be infected if vaccinated but uninfected if unvaccinated.
Under Assumption~\ref{ass1}, there are three principal strata or subgroups of
households defined by the joint potential infection outcomes of individual $%
1 $ under vaccine and control. They are (i) the doomed principal stratum in
which individual $1$ is infected whether vaccinated or not, (ii) the
protected stratum in which individual $1$ is infected if unvaccinated and
uninfected if vaccinated, and (iii) the immune stratum, in which individual $%
1$ does not become infected whether vaccinated or not. The causal contrast
(\ref{equ7}) is defined in the doomed principal stratum.

To simplify notation, let $p_{v}=E[Y_{i2}(1,0)|\break Y_{i1}(1,  0)=Y_{i1}(0,0)=1]$, $%
p_{u}=E[Y_{i2}(0,0)|Y_{i1}(1,0)=  Y_{i1}(0,0)=1]$, $%
p_{1}=E[Y_{i2}|Z_{i1}=1,Y_{i1}=1]$ and $p_{0}=E[Y_{i2}|Z_{i1}=0,Y_{i1}=1]$.
The crude (net) infectiousness effect (\ref{equ6}) is then just $p_{1}-p_{0}$, and
the causal infectiousness effect (\ref{equ7}) is $p_{v}-p_{u}$. Under randomization
and monotonicity, any household where individual 1 is infected if vaccinated
must be in the doomed stratum, so $p_{v}=p_{1}$. Thus, one component of the
causal infectiousness effect (\ref{equ7}) is identified.

However, any household where individual 1 becomes infected if unvaccinated
could be in the doomed or protected stratum. Thus, $p_{u}$ is not identified
without further assumptions. However, under monotonicity, the ratio $\rho$
of the proportion in the protected stratum to the sum of the proportions in
the protected and doomed strata is identified by the observed data.
Thus, we know what proportion of the households in which individual 1
received control and was infected is in the doomed stratum, just not which
ones, so we do not know what proportion of secondary transmissions occurred
in the doomed strata. Under Assumption~\ref{ass1}, Halloran and Hudgens (\citeyear{7},
\citeyear{8}) derived upper and lower bounds for causal effects on infectiousness
that are constrained by the relation in the data between $\rho $ and $p_{0}$.

A further possible assumption is the following:

\begin{assumption}\label{ass2}
$E[Y_{i2}(0,0)|Z_{i1}=0,Y_{i1}=1]\leq
E[Y_{i2}(0,0)|Z_{i1}=1,Y_{i1}=1]$.
\end{assumption}

 Assumption~\ref{ass2} states that the average infection rate for individual
$2$ if both individuals $1$ and $2$ were unvaccinated would be lower in the
subgroup of households for which individual $1$ would be infected and
unvaccinated than in the subgroup of households for which individual $1$
would be infected and vaccinated. The assumption might be thought plausible
insofar as the subgroup for which individual $1$ was vaccinated and infected
might be less healthy than the subgroup for which individual $1$ was
unvaccinated and infected; thus, if both people are unvaccinated, individual
$2$ is more likely to be infected in the first subgroup than in the second.

Under Assumptions \ref{ass1} and \ref{ass2}, \citet{41} showed
the crude contrast in (\ref{equ6}) is conservative for the causal contrast in (\ref{equ7}) in
that $E[Y_{i2}(1,0)-Y_{i2}(0,0)|Y_{i1}(1,0)=Y_{i1}(0,0)=1]\leq
E[Y_{i2}|Z_{i1}=1,Y_{i1}=1]-E[Y_{i2}|Z_{i1}=0,Y_{i1}=1] $, that is, $%
p_{v}-p_{u}\leq p_{1}-p_{0}$. Analogous results in fact also hold for the
risk ratio, odds ratio and vaccine efficacy scales (\cite{41}).


The approach may be employed outside of the vaccine context.  For example,
in an observational study in which the treatment is a smoking cessation
program in which one of two persons in a household participated. The
participation of the first person might affect the smoking behavior of the
second. This might occur either (i) because smoking cessation for the first
person encourages the second to stop smoking or because (ii)~even if the
first person does not stop smoking, the second person might nevertheless be
exposed to some of the smoking cessation program materials. One could
evaluate this second type of effect (the analogue of the infectiousness
effect) by applying the approach described above.

\section{Interference and Sensitivity Analysis}\label{sec3}

\subsection{Overview}\label{sec3.1}

In this section we develop sensitivity analysis techniques that can help
assess the presence of causal effects in two settings where these effects in
the presence of interference are not identified. These causal effects may
not be identified either because the treatments are not randomized or
because, even if the treatments are randomized, the spillover effects of
interest involve conditioning on a post-treatment variable thereby breaking
randomization. Building on the previous sections, we first consider the
setting of a randomized trial where the spillover effect of interest is not
identified by randomization alone because of conditioning on a
post-randomization variable as in the infectiousness effect described in
Section~\ref{sec2.6}. We present sensitivity analysis methods for assessing this
infectiousness effect in part by adapting research in
\citet{24}. We then consider the setting of observational data
such as in \citet{11} in which causal effects and spillover
effects may not be identified due to one or more unmeasured confounding
variables. We present two sensitivity analysis techniques for causal effects
in the presence of interference that extend analogous results for causal
effects under no-interference (\cite*{24}; \cite{39}) to the setting of causal effects and spillover effects in the presence
of interference.

\subsection{Sensitivity Analysis for the Infectiousness Effect}\label{sec3.2}

In Section~\ref{sec2.6} we described two previously developed approaches to bounds
on the infectiousness \mbox{effects}. Here we develop methods for sensitivity
analysis for the infectiousness effect. We follow the development first in
\citet{41} and \citet{7}
and then in \citet{12}; further technical development is
given in the \hyperref[app]{Appendix}. See also \citet{41}
and \citet{12} for concrete applications. A simple
sensitivity analysis approach also follows from the development of
\citet{41}. We use the same notation as in
Section~\ref{sec2.6}. As noted in Section~\ref{sec2.6}, under monotonicity Assumption~\ref{ass1}, we
have that $p_{v}=p_{1}$ and, thus, to obtain the causal infectiousness effect,
we need to express $p_{u}$ in terms of the observed data and sensitivity
analysis parameters. We will describe three different parameterizations.

First, let $\theta
=E[Y_{i2}(0,0)|Z_{i1}=1,Y_{i1}=1]-\break  E[Y_{i2}(0,0)|Z_{i1}=0,Y_{i1}=1]$ denote
the sensitivity parameter which contrasts the average counterfactual
infection rates for individual $2$ if both individuals $1$ and $2$ were
unvaccinated in the subgroup of households for which individual $1$ is
vaccinated and infected versus the subgroup of households for which
individual $1$ is unvaccinated and infected. It follows from the development
in VanderWeele and Tchetgen Tchetgen that, under monotonicity, $%
p_{u}=p_{0}+\theta $ and, thus,
\[
p_{v}-p_{u}=p_{1}-p_{0}-\theta.
\]%
In other words, to obtain the infectiousness effect under monotonicity, we
can calculate the crude infectiousness effect in (\ref{equ6}), specify the
sensitivity parameter $\theta $ and subtract the sensitivity parameter $%
\theta $ from the crude estimate to obtain the infectiousness effect. We can
vary $\theta $ over a range of plausible values in a sensitivity analysis to
produce a range of plausible values for the infectiousness effect. The
sensitivity analysis parameter is subject to certain empirical constraints
as described below. However, because of the simple relationship above, a
corrected confidence interval under sensitivity parameter $\theta $ can be
obtained simply by subtracting $\theta $ from both limits of the confidence
interval for the crude estimate in~(\ref{equ6}).

We can also use a similar approach but with a different parameterization of
the sensitivity analysis parameters. Following \citet{12},
we can vary $\gamma =E[Y_{i2}(0,0)|Y_{i1}(1,0)=0,Y_{i1}(0,0)=1]$, the
probability of secondary transmission in the protected stratum when
individual 1 receives control with bounds set by constraints of the data
(Halloran and Hudgens, \citeyear{7}, \citeyear{8}). The quantity $\gamma $ is not
identifiable from the observed data without further assumptions, but once a
value of $\gamma $ is assumed, then the probability of secondary
transmission in the doomed stratum is fixed and, thus, $p_{u}$ is identified.
Varying $\gamma $, the infectiousness effect (\ref{equ7}) on the risk difference
scale can be obtained as $p_{1}-p_{u}$, where $p_{u}$ is given by
\[
p_{u}=\frac{p_{0}-\gamma (1-\rho )}{\rho } ,
\]%
and where $\gamma $ can vary between
\[
\max \biggl\{ 0,\frac{p_{0}-\rho }{1-\rho } \biggr\} \leq \gamma \leq \min \biggl\{ 1,
\frac{p_{0}}{1-\rho } \biggr\} ,
\]%
with the left side giving rise to the upper bound for the causal
infectiousness effect and the right side giving rise to the lower bound on
the risk difference scale. Note that a drawback of this approach is that the
parameter $\gamma $ is constrained by the observed data and similar
restrictions are imposed on $\theta$ above by virtue of the identity $%
\theta=p_{0}-\frac{p_{0}-\gamma (1-\rho )}{\rho }$.\vspace*{1pt}

As a third parameterization, we could follow an approach to sensitivity analysis
 developed in \citet{29} and \citet{24}.
This approach performs sensitivity analysis with a bias parameter $\beta$ on the
ratio scale.
\citet{6} and  \citet{12} adapted this approach for sensitivity analysis for causal effects on
 post-infection outcomes, the former in the continuous post-infection outcome scenario, the
 latter for binary post-infection outcomes.  \citet{12} and
 \citet{7} suggested that this approach to sensitivity
 analysis could be used  for infectiousness effects taking as the intermediate infection outcome the infection
status of individual 1 and as the potential post-infection outcome the
infection status of individual 2.
 Within the context of
the infectiousness effect, again under monotonicity Assumption \ref{ass1}, the bias
parameter $\beta$  can be expressed as
\begin{eqnarray*}
\hspace*{-4.5pt}&&\exp (\beta )\\
\hspace*{-4.5pt}&&\quad=%
\bigl(P\bigl(Y_{i2}(0,0)=1|Z_{i1}=1,Y_{i1}=1
\bigr)
\\
\hspace*{-4.5pt}&&\qquad\hspace*{4.5pt}/P\bigl(Y_{i2}(0,0)=0|Z_{i1}=1,Y_{i1}=1\bigr)
\bigr)
\\
\hspace*{-4.5pt}&&\qquad/\bigl(P\bigl\{Y_{i2}(0,0)=1|Y_{i1}(0,0)=1,Y_{i1}(1,0)=0
\bigr\}
\\
\hspace*{-4.5pt}&&\qquad\hspace*{9pt}/P\bigl\{Y_{i2}(0,0)=0|Y_{i1}(0,0)=1,Y_{i1}(1,0)=0
\bigr\}\bigr).
\end{eqnarray*}
The bias parameter $\beta $ is the log of odds ratio comparing the risk of
infection if individual 1 is not vaccinated among (i) the doomed stratum and
(ii) the protected. Note this is not simply a different scale than the bias
parameter $\theta $ above (odds ratio versus risk difference) but also a
comparison of different subpopulations. Once this bias parameter is
specified, then it can be shown\vspace*{1pt} that $%
p_{u}=E[Y_{i2}(0,0)|Y_{i1}(1,0)=Y_{i1}(0,0)=1]$ is the positive root of $({%
-b\pm \sqrt{b^{2}-4zc}})/{2z}$, where
\begin{eqnarray*}
z &=&\exp (\beta )p_{0},
\\
b &=& \biggl( 1-\frac{P(Y_{i1}=1|Z_{i1}=1)}{P(Y_{i1}=1|Z_{i1}=1)} \biggr) \bigl\{\exp (\beta )-1\bigr\}
\\
&&{}-\exp (\beta )p_{0}+p_{0}-\exp (\beta ) ,
\\
c &=&\frac{P(Y_{i1}=1|Z_{i1}=1)}{P(Y_{i1}=1|Z_{i1}=1)}\bigl\{\exp (\beta )-1\bigr\}.
\end{eqnarray*}
For further discussion of inference for this approach to sensitivity analysis
see \citet{12} and \citet{14}.

In each of the three parameterizations above, once we have obtained $%
p_{u}=E[Y_{i2}(0,0)|Y_{i1}(1,0)=Y_{i1}(0, 0)=1]$, we can obtain the
infectiousness effect on the difference, risk ratio, odds ratio or
infectiousness effect scales by $p_{v}-p_{u}$, $p_{v}/p_{u}$, $%
p_{v}(1-p_{u})/\{p_{u}(1-p_{v})\}$ and $1-p_{v}/p_{u}$, respectively.

We have focused here on the setting of a randomized trial, but the approach
is potentially applicable to observational studies as well if, conditional
on some set of covariates $C$, the treatment was jointly independent of the
counterfactual outcomes (i.e., effectively randomized within strata of $C$).
 The sensitivity analysis parameters would have to be conditional on $C$.

\subsection{Sensitivity Analysis for Spillover Effects Under
Unmeasured Confounding: Approach~1}\label{sec3.3}

We now consider a setting in which causal effects and spillover effects
under interference are not identified due to unmeasured confounding.
Adjustment is often made for covariates to attempt to control for such
confounding. However, in an observational study we can never be sure that
the control is adequate. One or more unmeasured confounders may bias effect
estimates. Confounding control becomes even more complex in settings with
interference since when one individual's outcome is under consideration,
control will often need to be made for the covariates of other individuals
in the same cluster (\cite{32}; \cite{20}; \cite{21}). Unmeasured confounding can
thus operate either through the unmeasured covariates for the focal
individual or for other individuals in the same cluster. In this
subsection, we apply and extend the sensitivity analysis approach of
\citet{39} to allow for settings with interference and
spillover effects. In the next subsection we consider an extension of the
sensitivity analysis approach of \citet{24} to allow for
interference and spillover effects.

We consider a general observational setting such as that employed by \citet{11} wherein individuals are clustered in groups such that
individuals within groups may influence one another but there is no
interference between groups. We make the stratified interference assumption
above and further assume, following Hong and Raudenbush, that the potential
outcome of person $j$, $Y_{ij} ( \mathbf{z}_{i} ) $, depends on the
treatment received by the individuals in cluster $i$ other than person $j$,
$\mathbf{z}_{i(j)}$, only through some known many-to-one scalar function $g(%
\mathbf{z}_{i(j)})$ so that $Y_{ij} ( \mathbf{z}_{i} ) $ can be
written as $Y_{ij} ( z_{ij},g(\mathbf{z}_{i(j)}) ) $. For example,
$g(\mathbf{z}_{i(j)})$ might be the mean of $\mathbf{z}_{i(j)}$. Let $%
G_{ij}=g(\mathbf{Z}_{i(j)})$. Suppose that for all $i,j$, $Z_{ij}$ is determined by
simple randomization. We then have that
\[
E\bigl[Y ( z,g ) |Z=z,G=g\bigr]=E\bigl[Y ( z,g ) \bigr].
\]%
\citet{11} considered a variation on this assumption in the
context of observational data. Specifically, for some covariate vector $%
L_{ij}$, they assumed that%
\begin{eqnarray}
\label{equ8} && E\bigl[Y ( z,g ) |Z=z,G=g,L=l\bigr]
\nonumber
\\[-8pt]
\\[-8pt]
&&\quad=E\bigl[Y ( z,g ) |L=l\bigr]
\nonumber
\end{eqnarray}%
and from this it follows that
\[
E\bigl[Y ( z,g ) |L=l\bigr]=E[Y|Z=z,G=g,L=l],
\]%
where the right-hand side can be estimated with observed data. \citet{11} also allowed $L_{ij}$ to contain cluster level covariates
along with cluster aggregates of individual level covariates. Note, however,
that (\ref{equ8}) requires that $Y_{ij} ( z,g ) $ be mean independent of
both $Z_{ij}$ and $g(\mathbf{Z}_{i(j)})$ conditional on $L_{ij}$. If, for
each individual, $Z_{ij}$ is randomized conditional on $L_{ij}$, although
this will imply that $Y_{ij} ( z,g ) $ is mean independent of $%
Z_{ij}$ conditional on $L_{ij}$, it does not necessarily guarantee that $%
Y_{ij} ( z,g ) $ is mean independent of $g(\mathbf{Z}_{i(j)})$
conditional on $L_{ij}$. Let $\mathbf{L}_{i(j)}$ denote the vector of all
covariates $L_{ij}$ for all individuals in cluster $i$ other than individual
$j$. We might, instead of (\ref{equ8}), consider
\begin{eqnarray}
\label{equ9} &&E\bigl[Y ( z,g ) |Z=z,G=g,L=l,h(\mathbf{L})=h\bigr]
\nonumber
\\[-8pt]
\\[-8pt]
&&\quad=E\bigl[Y ( z,g ) |L=l,h(\mathbf{L})=h\bigr],
\nonumber
\end{eqnarray}%
where $h(\mathbf{L}_{i(j)})$ is a known function of $\mathbf{L}_{i(j)}$.
However, once again, with (\ref{equ9}), even if, for each individual, $Z_{ij}$ were
randomized conditional on $L_{ij},h(\mathbf{L}_{i(j)})$, this would not
guarantee that $Y_{ij} ( z,g ) $ is mean independent of $g(\mathbf{Z%
}_{i(j)})$ conditional on $L_{ij},h(\mathbf{L}_{i(j)})$ unless $h(\mathbf{L}%
_{i(j)})=\mathbf{L}_{i(j)}$. See \citet{20} for
discussion of causal structures for which assumptions (\ref{equ8}) or (\ref{equ9}) will hold.
Under assumption (\ref{equ9}), we have
\begin{eqnarray*}
&& E\bigl[Y ( z,g ) |L=l,h(\mathbf{L})=h\bigr]
\\
&&\quad=E\bigl[Y|Z=z,G=g,L=l,h(\mathbf{L})=h\bigr],
\end{eqnarray*}%
where again the right-hand side can be estimated with observed data. From
this one could obtain conditional direct, indirect and total effects, namely,%
\begin{eqnarray*}
&& E\bigl[Y(z,g)|l,h\bigr]-E\bigl[Y\bigl(z^{\prime },g\bigr)|l,h\bigr],
\\
&& E\bigl[Y(z,g)|l,h\bigr]-E\bigl[Y\bigl(z,g^{\prime }\bigr)|l,h\bigr],
\\
&& E\bigl[Y(z,g)|l,h\bigr]-E\bigl[Y\bigl(z^{\prime },g^{\prime }
\bigr)|l,h\bigr].
\end{eqnarray*}
These contrasts are important insofar as they allow one to assess the
relative importance for an individual's outcome of changing an individual's
own treatment versus the treatment of other individuals. In other words, the
effects allow one to assess the relative importance of spillover. Marginal
effects, involving counterfactuals of the form $E[Y ( z,g ) ]$,
could be obtained by averaging over the distributions of $L_{ij}$ and $h(%
\mathbf{L}_{i(j)})$.

Suppose now that we have unmeasured confounding by one or more unmeasured
confounders $U_{ij}$ and let $\mathbf{U}_{i(j)}$ denote the vector of $%
U_{ij} $ for all individuals in cluster $i$ other than individual $j$.
Suppose that the analogue of assumption (\ref{equ9}) holds conditional on observed $%
L_{ij},h(\mathbf{L}_{i(j)})$ and unobserved $U_{ij},v(\mathbf{U}_{i(j)})$
for some scalar function $v$ so that%
\begin{eqnarray}
\label{equ10}
&& E\bigl[Y ( z,g ) |\nonumber\\
&&\qquad Z=z,G=g,L=l,h(\mathbf{L})=h,U,v(\mathbf{U})
\bigr]
\\
&&\quad=E\bigl[Y ( z,g ) |L=l,h(\mathbf{L})=h,U,v(\mathbf{U})\bigr],
\nonumber
\end{eqnarray}%
but that (\ref{equ9}) does not hold when we do not condition on $U_{ij},v(\mathbf{U}%
_{i(j)})$. Without data on $U_{ij}$ causal effects are not identified. Let $%
H=h(\mathbf{L}_{i(j)})$ and $V=v(\mathbf{U}_{i(j)})$.

Following the sensitivity analysis approach of  \citet{39}
for causal effects under no-interference, we express the difference between
the causal effect
\begin{eqnarray*}
&&E\bigl[Y(z,g)|l,h\bigr]-E\bigl[Y\bigl(z^{\prime },g^{\prime
}
\bigr)|l,h\bigr]
\\
&&\quad=\sum_{u,v}\bigl\{E[Y|z,g,l,h,u,v]\\
&&\hspace*{28pt}\quad{}-E
\bigl[Y|z^{\prime },g^{\prime
},l,h,u,v\bigr]\bigr\}
\\
&&\hspace*{22pt}\quad{}\cdot P(u,v|l,h)
\end{eqnarray*}
and the biased estimand
\[
E[Y|z,g,l,h]-E\bigl[Y|z^{\prime },g^{\prime },l,h\bigr]
\]%
in terms of sensitivity analysis parameters. Let $B=\{E[Y|z,g,l,h]-E[Y|z^{%
\prime },g^{\prime },l,h]\}-\{E[Y(z,g)|l,h]-E[Y(z^{\prime },g^{\prime
})|l,h]\}$ denote this difference. Technical development is given in the
\hyperref[app]{Appendix}.

Let $u^{\ast }$ and $v^{\ast }$ denote arbitrary reference values for $U$
and $V$, respectively. Under assumption (\ref{equ10}) we have that%
\begin{eqnarray}
\label{equ11} B &=&\sum_{u,v}\bigl
\{E(Y|z,g,l,h,u,v)-E\bigl(Y|z,g,l,h,u^{\ast },v^{\ast
}\bigr)\bigr\}
\nonumber
\\
&&\hspace*{10pt} {}\cdot\bigl\{P(u,v|z,g,l,h)-P(u,v|l,h)\bigr\} 
\nonumber\\
&&{}-\sum_{u,v}\bigl\{E\bigl(Y|z^{\prime },g^{\prime },l,h,u,v
\bigr)\\
&&\hspace*{28pt} {}-E\bigl(Y|z^{\prime
},g^{\prime },l,h,u^{\ast },v^{\ast }
\bigr)\bigr\}
\nonumber
\\
&&\hspace*{23pt} {}\cdot\bigl\{P\bigl(u,v|z^{\prime },g^{\prime
},l,h
\bigr)-P(u,v|l,h)\bigr\}.
\nonumber
\end{eqnarray}%
To obtain the bias factor $B$, one could thus specify the effect of the
unmeasured confounders $U$ and $V$ on the outcome, $%
E(Y|z,g,l,h,u,v)-E(Y|z,g,l,h,\break u^{\ast },v^{\ast })$, for $(Z,G)=(z,g)$ and $%
(Z,G)=(z^{\prime },g^{\prime })$, and also how the distribution of $U$ and $%
V $ differs when $(Z,G)=(z,g)$ versus $(Z,G)=(z^{\prime },g^{\prime })$,
that is, $P(u,v|z,g,l,h)$ and $P(u,v|z^{\prime },g^{\prime },l,h)$. One can use
these sensitivity analysis parameters to calculate the bias factor in (\ref{equ11})
and then subtract the bias factor $B$ from the estimate of the causal effect
using the observed data $E[Y|z,g,l,h]-E[Y|z^{\prime },g^{\prime },l,h]$ to
obtain a corrected effect estimate for $E[Y(z,g)|l,h]-E[Y(z^{\prime
},g^{\prime })|l,h]$.

Note that the expression for the bias factor in (\ref{equ11}) makes no assumption
beyond assumption (\ref{equ10}) that control for observed $(L,H)$ and unobserved $%
(U,V)$ would suffice to control for confounding of the effect of $(Z,G)$ on $%
Y$; it allows for multiple unmeasured confounders. However, the use of the
bias formula in (\ref{equ11}) requires specifying a large number of parameters: $%
E(Y|z,g,l,h,u,v)-E(Y|z,g,l,h,u^{\ast },v^{\ast })$ for every value of $u,v$
and the distributions $P(u,v|z,g,l,h)$ and $P(u,v|z^{\prime },g^{\prime
},l,h)$.

Under some simplifying assumptions, expression (\ref{equ11}) reduces to a much easier
to use formula. In particular, suppose that there is a single unmeasured
confounder $U$ and that $V=v(\mathbf{U}_{i(j)})$ is scalar. Suppose also
that the effects of $U$ and $V=v(\mathbf{U}_{i(j)})$ are additive in the
sense that $E(Y|z,g,l,h,u,v)-E(Y|z,g,l,h,u^{\ast },v^{\ast })=\lambda
(u-u^{\ast })+\tau (v-v^{\ast })$ for $(Z,G)=(z,g)$ and $(Z,G)=(z^{\prime
},g^{\prime })$. In the \hyperref[app]{Appendix} it is shown that under these assumptions%
\begin{eqnarray}
\label{equ12} \hspace*{15pt}B&=&\lambda \bigl\{E[U|z,g,l,h]-E
\bigl[U|z^{\prime },g^{\prime },l,h\bigr]\bigr\}
\nonumber
\\[-8pt]
\\[-8pt]
\hspace*{15pt}&&{}+\tau \bigl\{E[V|z,g,l,h]-E\bigl[V|z^{\prime },g^{\prime },l,h
\bigr]\bigr\}.
\nonumber
\end{eqnarray}%
To use this simplified bias formula, one only needs to specify the effect, $%
\lambda $, for a one unit increase in the unmeasured confounder $U_{ij}$,
the effect $\tau $ of a one unit increase in the scalar functional of the
unmeasured confounders of the other members of the group, $v(\mathbf{U}%
_{i(j)})$, and how the means of $U_{ij}$ and $v(\mathbf{U}_{i(j)})$ differ
when $(Z,G)=(z,g)$ versus when $(Z,G)=(z^{\prime },g^{\prime })$. Once these
sensitivity analysis parameters are specified the bias factor $B$ can be
calculated using formula (\ref{equ12}) above and then $B$ could be subtracted from
the estimate of the causal effect using the observed data $%
E[Y|z,g,l,h]-E[Y|z^{\prime },g^{\prime },l,h]$ to obtain a corrected effect
estimate for $E[Y(z,g)|l,h]-E[Y(z^{\prime },g^{\prime })|l,h]$. Under this
simplified approach because the bias factor involves only the sensitivity
analysis parameters and not the observed data, a corrected confidence
interval could be obtained by subtracting $B$ from both limits of a
confidence interval for $E[Y|z,g,l,h]-E[Y|z^{\prime },g^{\prime },l,h]$. A
sensitivity analysis consists of reporting the causal contrasts under a
range of plausible values of $\tau$ and $\lambda$. The values $\tau=\lambda
= 0$ correspond to the assumption of no unmeasured confounders. Selecting a
plausible range of parameters could be done using subject matter expertise,
by external data from other studies or by including a very wide range of
parameters that are thought to include those that would constitute very
extreme values. Alternatively, one could examine the most important measured
confounder and assess the magnitude of the corresponding parameters for the
most important measured confounder; one could then consider whether an
additional unmeasured confounder with parameters set equal to that of the
most important measured confounder would substantially alter results. This
would allow an investigator to assess whether an unmeasured confounder would
have to be stronger than the most important measured confounder to
substantially alter the results.

Consider again the substantive example of \citet{11}
described in Section~\ref{sec1.1.3}. As noted above, \citet{11}
examined the effect of kindergarten retention on reading test scores
allowing for interference by allowing the retention of other students at the
school to affect a child's reading test scores. They assumed that treatment
assignment at both the school and the individual level was ignorable given a
number of observed individual-level, school-level and
school-aggregated-individual-level characteristics. Using a propensity score-based approach, they estimated, in the notation above, the contrasts $%
E[Y(z=1,g=1)]-E[Y(z=0,g=1)]$ and $E[Y(z=1,g=0)]-E[Y(z=0,g=0)]$, where $g=1$
denotes high-retention school and $g=0$ a low-retention school. They found
that, in low-retention schools, their estimates indicated that the effect on
reading scores of a student being retained versus being promoted was $-8.18$
(95\% CI: $-$10.02, $-$6.34), and in high retention schools the effect estimate
was $-8.86$ (95\% CI: $-$11.56, $-$6.16). A standard deviation in reading test
scores in this sample is 13.48 points.

Hong and Raudenbush also went through a sensitivity analysis argument for
their results. They noted that the strongest predictor of current test
scores were lagged test scores, but that it was unlikely that there was any
unmeasured covariate that would predict their outcomes so strongly. They
considered instead whether unmeasured individual and school covariates that
had effects on readings scores that were equal to those of the measured
covariates with second strongest association with reading scores would
suffice to explain away the effect estimates. Using an argument based on a
formula similar to (\ref{equ12}), they reported that unmeasured individual and school
confounders that had an effect as large as the second most important
measured individual and school level covariates would shift the estimate in
high-retention schools to $-4.25$ (95\% CI: $-$6.95, $-$1.54) and thus not
suffice to bring the confidence interval to include $0$. However, in
low-retention schools, unmeasured individual and school confounders that had
an effect as large as the second most important measured covariates would
shift the  estimate and confidence  interval in low-retention schools to $-0.60$ (95\%
CI: $-$2.44, 1.24) and thus would suffice to bring their confidence
interval for the effect in low-retention schools to include  $0$. The effects in
high-retention schools seem more robust to the possibility of unmeasured
confounding. In their analyses, \citet{11} used a similar
expression to (\ref{equ12}), but in their paper they did not provide a derivation of
this formula and did not articulate the assumptions needed for the use of
the formula. We have provided the derivations and assumptions required here.
Moreover, we have also provided a more general expression, that in (\ref{equ11}),
that is applicable under much weaker assumptions.

We have considered here a sensitivity analysis approach for causal effects
and spillover effects in the presence of interference. In other contexts,
questions concerning whether the effect of a treatment on an outcome is
mediated by some intermediate may be of interest. In settings in which
mediation is of interest and interference occurs at the level of the
mediator so that the mediator for one unit may affect the outcomes for other
units (cf. \cite*{36}; VanderWeele et al., \citeyear{43}), a similar
sensitivity analysis approach for unmeasured confounding of the mediator and
the outcome could be developed by applying and extending the results of
\citet{37} for direct and indirect effects from the no-interference
setting to a setting with interference by following an analogous approach to
that presented above.

\subsection{Sensitivity Analysis for Spillover Effects
Under Unmeasured Confounding: Approach~2}\label{sec3.4}

In this subsection we consider an alternative sensitivity analysis approach
to assess the influence of unobserved confounding for direct and spillover
effects in general settings similar to those considered by \citet{11} described above.
The following developments follow closely from analogous sensitivity analysis techniques recently
proposed in the context of mediation analysis (\cite{33}; \cite{34}). In order to formalize the approach,
suppose that we wish to make inferences about the following causal effects:
\begin{eqnarray*}
\gamma ^{d} ( z,g,l,h ) &=&E\bigl[Y ( z,g ) -Y ( z_{0},g )
|z,g,l,h\bigr],
\\
\gamma ^{s} ( g,l,h ) &=&E\bigl[Y ( z_{0},g ) -Y (
z_{0},g_{0} ) |g,l,h\bigr],
\end{eqnarray*}%
where, unless stated otherwise, throughout $G$ and $H$ are left
unrestricted, that is, $G_{i(j)}=g(\mathbf{Z}_{i(j)})=\mathbf{Z}%
_{i(j)},H_{i(j)}=h(\mathbf{L}_{i(j)})=\mathbf{L}_{i(j)}$. These effects are
versions, allowing for interference, of the so-called effect of treatment on
the treated, which have been studied extensively in the absence of
interference by econometricians, epidemiologists and social scientists. The
first contrast $\gamma ^{d} ( z,g,l,h ) $ captures the direct
effect of $Z_{ij}$ on $Y_{ij}$ conditional on the person's observed exposure
$Z_{ij}$ and the cluster's observed data $(\mathbf{Z}%
_{i(j)},L_{ij},H_{i(j)})$. In contrast, $\gamma ^{s} ( g,l,h ) $
is the spillover causal effect\vspace*{2pt} of $\mathbf{Z}_{i(j)}$ on $Y_{ij} (
z_{i,j}=z_{0} ) $ within levels of $(\mathbf{Z}_{i(j)},L_{ij},H_{i(j)})$.
Note that $\gamma ^{d} ( z_{0},g,l,h ) =\gamma ^{s} (
g_{0},l,h ) =0$, so that these effects are relative to the reference
average potential outcome under $ ( z_{0},g_{0} ) $.

Consider the pair of no unobserved confounding assumptions:
\begin{eqnarray*}
E\bigl[Y ( z_{0},g ) |z,g,l,h\bigr] &=&E\bigl[Y ( z_{0},g
) |z_{0},g,l,h\bigr],
\\
E\bigl[Y ( z_{0},g_{0} ) |g,l,h\bigr] &=&E\bigl[Y (
z_{0},g_{0} ) |g_{0},l,h\bigr].
\end{eqnarray*}

 It is straightforward to show that under these assumptions, $\gamma ^{d}$
and $\gamma ^{s}$ are identified by
\begin{eqnarray*}
\gamma ^{d,\dag } ( z,g,l,h ) &=&E[Y|z,g,l,h]
\\
&&{} -E[Y|z_{0},g,l,h],
\\
\gamma ^{s,\dag } ( g,l,h ) &=&E[Y|z_{0},g,l,h]
\\
&{}&-E[Y|z_{0},g_{0},l,h].
\end{eqnarray*}%
This shows that the above no unobserved confounding assumption suffices to
identify direct and spillover effects (on the treated) using a standard
regression analysis approach to model $E[Y|z,g,l,h]$. Next, supposing that the
no unobserved confounding assumption does not hold, we define the following
selection bias functions:
\begin{eqnarray}
\label{equ13} \delta ^{d} ( z,g,l,h ) &=&E\bigl[Y (
z_{0},g ) |z,g,l,h\bigr]
\nonumber
\\[-8pt]
\\[-8pt]
&&{}-E\bigl[Y ( z_{0},g ) |z_{0},g,l,h\bigr],
\nonumber
\\
\label{equ14} \delta ^{s} ( g,l,h ) &=&E\bigl[Y (
z_{0},g_{0} ) |g,l,h\bigr]
\nonumber
\\[-8pt]
\\[-8pt]
&&{}-E\bigl[Y ( z_{0},g_{0} ) |g_{0},l,h\bigr],
\nonumber
\end{eqnarray}%
where $\delta ^{d} ( z_{0},g,l,h ) =\delta ^{s} (
g_{0},l,h ) =0$. These selection bias functions come about naturally
upon contrasting, on the additive scale, each of the observational
conditional association $\gamma ^{d,\dag } ( z,g,l,h ) $ and $%
\gamma ^{s,\dag } ( g,l,h ) $, with their corresponding causal
analog, $\gamma ^{d} ( z,g,l,h ) $ and $\gamma ^{s} (
g,l,h ) $, respectively. To illustrate in the simple context of binary $%
Z$, one can verify that the confounding bias quantified on the
additive scale is equal to
\begin{eqnarray*}
&&\gamma ^{d,\dag } ( z=1,g,l,h ) -\gamma ^{d} ( z=1,g,l,h )
\\
&&\quad=E[Y|z=1,g,l,h]-E[Y|z_{0},g,l,h]
\\
&&\qquad{}-E \bigl[ Y ( z=1,g ) |z=1,g,l,h \bigr]
\\
&&\qquad{}+E \bigl[ Y ( z_{0},g ) |z=1,g,l,h \bigr]
\\
&&\quad=E\bigl[Y ( z=1,g ) |z=1,g,l,h\bigr]
\\
&&\qquad{}-E\bigl[Y ( z_{0},g ) |z_{0},g,l,h\bigr]
\\
&&\qquad{}-E \bigl[ Y ( z=1,g ) |z=1,g,l,h \bigr]
\\
&&\qquad{}+E \bigl[ Y ( z_{0},g ) |z=1,g,l,h \bigr]
\\
&&\quad=E \bigl[ Y ( z_{0},g ) |z=1,g,l,h \bigr]
\\
&&\qquad{}-E\bigl[Y ( z_{0},g ) |z_{0},g,l,h\bigr]
\\
&&\quad=\delta ^{d} ( z=1,g,l,h ),
\end{eqnarray*}%
which makes clear the central role of the selection bias function $\delta
^{d}$. A generalization of the above derivation gives similar expressions
for $\delta ^{s} ( g,l,h ) $ and also extends beyond binary $Z$.
Furthermore, this derivation also makes clear that the presence of
confounding implies that at least one of the following must hold:
\begin{eqnarray*}
\mbox{either}\quad\delta ^{d} ( z,g,l,h ) &\neq &0\quad\mbox{for some
}%
 ( z,g,l,h )
\\
\mbox{or}\quad\delta ^{s} ( g,l,h ) &\neq &0\quad\mbox{for some } (
g,l,h ) .
\end{eqnarray*}%
The first condition implies $\gamma ^{d}$ is not identified, while the
second case implies $\gamma ^{s}$ is not identified. Thus, we may proceed as
in \citet{24} and recover causal inferences by assuming the
selection bias functions, $\delta ^{d} ( z,g,l,h ) $ and $\delta
^{s} ( g,l,h ) $, that encode the magnitude and direction of
unmeasured confounding, are known. Suppose that higher values of $Y$ are
beneficial to one's health. If $\delta ^{d} (
1,g,l,h ) >0$, then, on average, an individual $j$ in cluster $i$ with $%
 \{ G_{i(j)}=g,L_{ij}=l,H_{i(j)}=h \} $ and exposure value $Z_{ij}=1
$ has higher potential outcomes $Y_{ij} ( z_{0}=0,g ) $ than an
individual in the same cluster and the same stratum $ \{
G_{i(j)}=g,L_{ij}=l,H_{i(j)}=h \} $ but unexposed $Z_{ij}=0$, that is,
healthier individuals are more likely to receive the exposure conditional on
the exposures of other people in the cluster and the observed confounders
for the cluster.  On the other hand, $\delta ^{d} ( 1,g,l,h ) <0$
suggests confounding by indication for exposure, that is, unhealthier
individuals are more likely to be exposed. Likewise, if $\delta ^{s} (
g,l,h ) >0$ for all $g\neq g_{0}$ indicates that, on average, an
individual in a cluster with confounders $ \{
L_{ij}=l,H_{i(j)}=h \} $ and $g(\mathbf{Z}_{i(j)})=g^{\ast }\neq g_{0}$
has higher potential outcomes $Y_{ij} ( z_{0}=0,g_{0} ) $ than a
comparable individual in a comparable cluster with baseline exposure value $%
g(\mathbf{Z}_{i(j)})=g_{0}$. In the special case where $g ( \mathbf{Z%
}_{i(j)} ) =\mathbf{Z}_{i(j)}=\mathbf{0,}$ one has that clusters with
no exposed individual tend, on average, to be less healthy than clusters with
one or more individuals exposed.

The approach to inference in the presence of confounding involves the
following reparameterization of the conditional mean function $E[Y|z,g,l,h]$
in terms of the causal contrasts $\gamma ^{d}$ and $\gamma ^{s}$ and the
selection bias functions $\delta ^{d}$ and $\delta ^{s}$. To state the
reparameterization, suppose for the moment that $f(z,g|l,h)=f ( \mathbf{Z%
}_{i}= ( z,g ) |\mathbf{L}_{i}= ( l,h )  ) $ is
known, then one can verify that for each unit in cluster $i$,
\begin{eqnarray*}
&&E[Y|z,g,l,h]
\\
&&\quad=\gamma ^{d} ( z,g,l,h ) +\delta ^{d} ( z,g,l,h )
\\
&&\qquad{} -\sum_{z=0}^{1}\delta
^{d} ( z,g,l,h ) f ( z|g,l,h )
\\
&& \qquad{}+\gamma ^{s} ( g,l,h ) +\delta ^{s} ( g,l,h )
\\
&&\qquad{}-\sum_{\mathbf{z}^{\ast }\in \{0,1\} ^{n_{i}-1}}\delta ^{s} \bigl( g
\bigl(\mathbf{z}^{\ast }\bigr),l,h \bigr) f \bigl( \mathbf{z}^{\ast }|l,h
\bigr) +q(l,h),
\end{eqnarray*}%
where
\[
q(l,h)\equiv E\bigl[Y ( z_{0},g_{0} ) |l,h\bigr].
\]%
For fixed $\delta ^{d}$ and $\delta ^{s}$ given by equations (\ref{equ13}) and (\ref{equ14}),
the causal contrasts $\gamma ^{d}$ and $\gamma ^{s}$ are nonparametrically
identified and can be estimated by fitting the above regression model.

In practice, due to the high dimensionality of $\mathbf{L}_{i}$ often
encountered in applications, parametric models must be used to reliably
estimate $\gamma ^{d} ( z,g,l,h ) ,\break \gamma ^{s} ( g,l,  h )
, f(z,g|l,h)$ and $q(l,h)$. The above reparameterization is particularly
advantageous in that it ensures variation independence of parameters of
working models for these various quantities. A description of
parametric maximum likelihood and generalized estimating equations
estimation is given in the \hyperref[app]{Appendix}.

\subsection{A Comparison of Sensitivity Analysis Techniques}\label{sec3.5}

It is instructive to compare the sensitivity analysis techniques given in
Sections~\ref{sec3.3} and \ref{sec3.4}. We begin by noting that the causal estimand targeted
by the two methods differ. The first approach aims to make inferences about
\begin{eqnarray*}
&&E\bigl[Y(z,g)-Y\bigl(z^{\prime },g\bigr)|l,h\bigr]\quad\mbox{and}
\\
&& E\bigl[Y\bigl(z^{\prime },g\bigr)-Y\bigl(z^{\prime
},g^{\prime }
\bigr)|l,h\bigr],
\end{eqnarray*}%
while the second approach targets the causal contrasts
\begin{eqnarray*}
&&E\bigl[Y(z,g)-Y\bigl(z^{\prime },g\bigr)|z,g,l,h\bigr]\quad\mbox{and}
\\
&&E\bigl[Y\bigl(z^{\prime },g\bigr)-Y\bigl(z^{\prime
},g^{\prime }
\bigr)|g,l,h\bigr].
\end{eqnarray*}%
In the absence of confounding, the contrasts targeted by the two approaches
coincide, but when unmeasured confounding is present, the first approach
gives direct and spillover effects for the subset of individuals with $%
 \{ l,h \} $, while the second approach delivers inferences about
the direct effect for individuals with $ \{ z,g,l,h \} $ and the
spillover effect for individuals with $ \{ g,l,h \} $, that is,
interference effects of treatment on the treated$.$ This distinction has
implications for the corresponding sensitivity analysis techniques. Although
both approaches require specifying unidentified parameters, in the first
technique the parameters correspond to particular causal effects (of $U$ and
$V$), whereas in the second technique the parameters do not correspond to
causal effects. More specifically, in the first approach, a parametrization
of the bias expression (\ref{equ11}) involves quantifying the causal interaction
\begin{eqnarray*}
&& \bigl\{ E(Y|z,g,l,h,u,v)-E\bigl(Y|z,g,l,h,u^{\ast },v^{\ast }
\bigr) \bigr\}
\\
&&\qquad{}- \bigl\{ E\bigl(Y|z^{\prime },g^{\prime },l,h,u,v\bigr)\\
&&\hspace*{38pt}{}-E
\bigl(Y|z^{\prime },g^{\prime },l,h,u^{\ast
},v^{\ast }
\bigr) \bigr\}
\\
&&\quad= \bigl\{ E(Y|z,g,l,h,u,v)-E\bigl(Y|z^{\prime },g^{\prime },l,h,u,v
\bigr) \bigr\}
\\
&&\qquad{}- \bigl\{ E\bigl(Y|z,g,l,h,u^{\ast },v^{\ast }\bigr)\\
&&\hspace*{38pt}{}-E
\bigl(Y|z^{\prime },g^{\prime
},l,h,u^{\ast },v^{\ast }
\bigr) \bigr\}
\end{eqnarray*}%
of the effect of $ ( Z,G ) $ within levels of $(L,H,U,V)$. The
first sensitivity analysis approach requires  making a judgement
   not only about the nature of $U$ (i.e., binary, polytomous, multivariate, etc.), and
the magnitude and the direction of unmeasured confounding, but also about
the magnitude and direction of effect heterogeneity on the additive scale.
In contrast, the second sensitivity analysis technique directly quantifies
the magnitude and direction of unmeasured confounding without making any
reference to a specific $U$ and, therefore, it does not involve making any
judgement about unidentified causal effects. While making a judgement about
the nature of $U$ (i.e., binary, polytomous, multivariate, etc.) in the first
approach could be difficult in practice, the second approach, to ensure that
posited models are compatible, requires that the user posit parametric
models for $f ( z|g,l,h ) $ and $f ( \mathbf{G}=g|l,h ) $,
an additional modeling requirement not needed by the first approach. Both of
these densities are, however, nonparametrically identified from the observed
data and, therefore, standard goodness-of-fit tools may be adopted to ensure
a reasonable fit to the data. The first approach can be of particular use if
subject matter expertise can help determine the nature of the unmeasured
confounders and/or if prior analyses with other data for which these
confounders are available can be used to help inform the value of the
sensitivity analysis parameters.

The difference between the two techniques also has interesting but subtle
implications if these techniques are used to construct tests of the sharp
null of no treatment effects. If an investigator were interested in testing
the sharp null of no treatment effects using the first sensitivity analysis
approach, care would need to be taken to ensure that the specification of
the sensitivity analysis parameters was compatible with the sharp null, for example,
by ensuring the interaction function in the above display is $0$ as, for
example, in the simplified expression in (\ref{equ12}). In contrast, with the second
sensitivity analysis approach, any specification of the sensitivity analysis
parameters will be compatible with the sharp null, and so this issue of
compatibility is not similarly a concern.

\section{Discussion}\label{sec4}

In this paper we have reviewed definitions and approaches to causal
inference in the presence of interference. We have developed various
sensitivity analysis approaches when the causal effects and spillover
effects of interest are unidentified either because randomization does not
suffice for identification (\cite*{12}; \cite{41}; Halloran and Hudgens, \citeyear{7}, \citeyear{8}) or because of
unmeasured confounding in an observational study.

We have extended existing sensitivity analysis approaches (\cite*{24}; \cite{39}) from the setting of no-interference to
allow for interference. Many settings in the social sciences in which causal
effects and spillover effects are of interest are observational settings
with interference and the results presented here will likely be useful in
those settings. Further work could be done on sensitivity analysis for
unmeasured confounding in other settings in which there are not multiple
independent clusters (e.g., \cite*{2}) or in settings involving
the assessment of causal effects in social networks (\cite{3}; \cite{38}).

More generally, numerous further challenges remain in the development of
methods for causal inference under interference. Inference may become
additionally challenging for treatment with multiple levels, as the number of
combinations will increase dramatically. Interference patterns might also
depend on the covariates of individuals in a cluster in complex ways.
Furthermore, the approach for defining causal estimands for infectiousness
where the covariates of individuals in the group are taken into account
becomes unwieldly when the clusters are larger than two, posing an
additional challenge for future research. When dealing with
cluster-randomized studies in which the clusters are large, the number of
clusters may be small, making inference difficult. In contrast, in the
household studies, the number of clusters may be large, but the number in
the households small.

One of the limitations of the approaches for causal inference with
interference presented in this paper is the assumption that there were fixed
groups or blocks of individuals. One of the challenges for future research
will be the issue of interference across groups. More generally, recent
research on causal inference with interference has been relaxing the
assumption of fixed groups. \citet{2} presented
randomization-based methods for estimating average causal effects under
arbitrary interference of known form. \citet{35} incorporates
network information for each individual under consideration that describes
the set of other individuals that each individual is potentially connected
to. Inference is then driven by the number of individuals rather than the number
of communities, which in this case is one. \citet{17} propose
new generalized inverse probability weighted estimators of causal effects in
the presence of any form of interference between individuals. Defining and
comparing causal estimands of effects across more general forms of
interference will be challenging. Much more exciting research is left to be
done on causal inference under general forms of interference.

\begin{appendix}\label{app}

\section*{Appendix}

\subsection*{Derivations for the Infectiousness Effect}\label{secA.1}

\citet{41} show\-ed that under monotonicity
assumption 1, $%
p_{v}=  E[Y_{i2}(1,0)|Y_{i1}(1,0)=Y_{i1}(0,0)=1]=E[Y_{i2}|Z_{i1}=1,Y_{i1}=1]=p_{1}
$ and $%
p_{u}=E[Y_{i2}(0,0)|Y_{i1}(1,0)=\break Y_{i1}(0,0)=1]=E[Y_{i2}|Z_{i1}=0,Y_{i1}=1]+%
\{E[Y_{i2}(0, 0)|\break Z_{i1}=1,Y_{i1}=1]-E[Y_{i2}(0,0)|Z_{i1}=0,Y_{i1}=1]\}=p_{1}+%
\theta $. From this it follows that $p_{v}-p_{u}= (p_{1}-p_{0})-\theta $ and $%
p_{v}/p_{u}=p_{1}/(p_{0}+\theta )$, $p_{v}(1-p_{u})/\{p_{u}(1-p_{v})%
\}=p_{1}(1-p_{0}-\theta )/\{(p_{0}+\theta )(1-p_{1})\}$, and $%
1-p_{v}/p_{u}=1-p_{1}/(p_{0}+\theta )$.

\citet{12} showed that under monotonicity assumption 1, $%
p_{u}=\gamma B/\{1+\gamma (B-1)\}$ and $p_{0}=\gamma V+p_{u}(1-V)$, where $%
B=\exp (\beta )$, $\gamma =P\{Y_{i2}(0,0)|Y_{i1}(1,0)=0,Y_{i1}(0,0)=1\}$,
and\vspace*{1pt} $V= ( 1-\frac{P(Y_{i1}=1|Z_{i1}=1)}{P(Y_{i1}=1|Z_{i1}=1)} ) $.
Solving $p_{u}=\gamma B/\{1+\gamma (B-1)\}$ and $p_{0}=\gamma V+p_{u}(1-V)$
to eliminate $\gamma $ gives $p_{u}=\frac{p_{0}-p_{u}(1-V)}{V}B/\{1+\frac{%
p_{0}-p_{u}(1-V)}{V}(B-1)\}$ or $%
0=p_{0}B-p_{u}(-V-Bp_{0}+p_{0}-B+VB)+(B-1)(1-V)p_{u}^{2}$, a quadratic
equation in $p_{u}$. The roots of this equation are $\frac{-b\pm \sqrt{%
b^{2}-4zc}}{2z}$, where $z=\exp (\beta )p_{0}$, $b= ( 1-\frac{%
P(Y_{i1}=1|Z_{i1}=1)}{P(Y_{i1}=1|Z_{i1}=1)} ) \{\exp (\beta )-1\}-\exp
(\beta )p_{0}+p_{0}-\exp (\beta )$, and $c=\frac{P(Y_{i1}=1|Z_{i1}=1)}{%
P(Y_{i1}=1|Z_{i1}=1)}\{\exp (\beta )-1\}$.

\subsection*{Derivations for Sensitivity Analysis of Spillover Effect
Under Unmeasured Confounding: Approach~1}\label{secA.2}

If in the notation of \citet{39}, we let $A$, $X$ and $U$
be $(Z,G)$, $(L,H)$ and $(U,V)$, respectively, and we let $a_{1}$ and $a_{0}$
denote $(z,g)$ and $(z^{\prime },g^{\prime })$, respectively, then by
\citet{39} we have that
\begin{eqnarray*}
B &=&E[Y|z,g,l,h]-E\bigl[Y|z^{\prime },g^{\prime
},l,h\bigr]\\
&&{}-E
\bigl[Y(z,g)|l,h\bigr]
-E\bigl[Y\bigl(z^{\prime },g^{\prime }\bigr)|l,h\bigr]
\\
&=&E[Y|z,g,l,h]-E\bigl[Y|z^{\prime },g^{\prime
},l,h\bigr]\\
&&{}-\sum_{u,v}\bigl\{E[Y|z,g,l,h,u,v]\\
&&\hspace*{30pt}{}-E
\bigl[Y|z^{\prime },g^{\prime
},l,h,u,v\bigr]\bigr\}
 P(u,v|l,h)
\\
&=&\sum_{u,v}\bigl\{E(Y|z,g,l,h,u,v)-E
\bigl(Y|z,g,l,h,u^{\ast },v^{\ast
}\bigr)\bigr\}
\\
&&\hspace*{11pt} {}\cdot\bigl\{P(u,v|z,g,l,h)-P(u,v|l,h)\bigr\}
\\
&&{}-\sum_{u,v}\bigl\{E\bigl(Y|z^{\prime },g^{\prime },l,h,u,v
\bigr)\\
&&\hspace*{30pt}{}-E\bigl(Y|z^{\prime
},g^{\prime },l,h,u^{\ast },v^{\ast }
\bigr)\bigr\}
\\
&&\hspace*{25pt} {}\cdot\bigl\{P\bigl(u,v|z^{\prime },g^{\prime
},l,h
\bigr)-P(u,v|l,h)\bigr\}.
\end{eqnarray*}%
If $E(Y|z,g,l,h,u,v)-E(Y|z,g,l,h,u^{\ast },v^{\ast })=\lambda (u-u^{\ast
})+\tau (v-v^{\ast })$, then we have
\begin{eqnarray*}
B &=&\sum_{u,v}\bigl\{E(Y|z,g,l,h,u,v)-E
\bigl(Y|z,g,l,h,u^{\ast },v^{\ast
}\bigr)\bigr\}
\\
&&\hspace*{11pt} {}\cdot\bigl\{P(u,v|z,g,l,h)-P(u,v|l,h)\bigr\}
\\
&&{}-\sum_{u,v}\bigl\{E\bigl(Y|z^{\prime },g^{\prime },l,h,u,v
\bigr)\\
&&\hspace*{28pt}{}-E\bigl(Y|z^{\prime
},g^{\prime },l,h,u^{\ast },v^{\ast }
\bigr)\bigr\}
\\
&&\hspace*{22pt} {}\cdot\bigl\{P\bigl(u,v|z^{\prime },g^{\prime
},l,h
\bigr)-P(u,v|l,h)\bigr\}
\\
&=&\sum_{u,v}\bigl\{\lambda \bigl(u-u^{\ast}
\bigr)+\tau \bigl(v-v^{\ast} \bigr)\bigr\}
\\
&&\hspace*{11pt} {}\cdot\bigl\{P(u,v|z,g,l,h)-P(u,v|l,h)\bigr\}
\\
&&{}-\sum_{u,v}\bigl\{\lambda \bigl(u-u^{\ast}
\bigr)+\tau \bigl(v-v^{\ast} \bigr)\bigr\}
\\
&&\hspace*{22pt} {}\cdot\bigl\{P\bigl(u,v|z^{\prime
},g^{\prime },l,h
\bigr)-P(u,v|l,h)\bigr\}
\\
&=&\sum_{u,v}(\lambda u+\tau v)P(u,v|z,g,l,h)
\\
&&{}-\sum_{u,v}(\lambda u+\tau v)P
\bigl(u,v|z^{\prime },g^{\prime },l,h\bigr)
\\
&=&\lambda \bigl\{E[U|z,g,l,h]-E\bigl[U|z^{\prime },g^{\prime },l,h
\bigr]\bigr\}
\\
&&{}+\tau \bigl\{E[V|z,g,l,h]-E\bigl[V|z^{\prime },g^{\prime },l,h\bigr]
\bigr\}.
\end{eqnarray*}

\subsection*{Maximum Likelihood and GEE Estimation of Direct and
Spillover Effects Under Unmeasured Confounding: Approach 2}\label{secA.3}

We briefly describe maximum likelihood and generalized estimating equations
inference for direct and spillover effects in the presence of unobserved
confounding under Approach 2. As stated in the text, in practice, due to the
high dimensionality of $\mathbf{L}_{i}$ often encountered in applications,
parametric models must be used to reliably estimate $\gamma ^{d} (
z,g,l,h ) ,\gamma ^{s} ( g,l,h ) ,  f ( \mathbf{Z}_{i}|%
\mathbf{L}_{i} ) $ and $q(l,h)$. The proposed reparameterization of $%
E[Y|z,g,l,h]$ is particularly advantageous in that it ensures variation
independence of parameters of the working models for these various
quantities, making possible a straightforward application of maximum
likelihood estimation.  Specifically, consider the parametric models $%
\gamma ^{d} ( z,g,l,h;\psi ^{d} ) ,\gamma ^{s} ( g,l,h;\psi
^{s} ) $, $f ( \mathbf{Z}_{i}|\mathbf{L}_{i};\alpha  ) $ and $%
q(l,h;\eta )$; then, provided parameters are not shared across working
models, a particular choice of one of these models is guaranteed by our
parametrization not to place any restriction on the other models. Maximum
likelihood estimation of unknown parameters requires that one posit an
additional working model for the conditional density $f(\boldsymbol{\varepsilon }%
_{i}\mathbf{|L}_{i})$ which we denote $f(\boldsymbol{\varepsilon }_{i}\mathbf{|L}%
_{i};\omega )$, for $\boldsymbol{\varepsilon }_{i}$ the vector of possibly
correlated residuals $Y_{ij}-$ $E[Y|z,g,l,h]$, $j=1,\ldots,n_{i}$. In
principle, our choice of parametrization could be used in conjunction with
standard techniques for modeling clustered outcomes, such as, for instance, by
incorporating a random intercept to introduce correlation within a cluster
in the regression model of $Y_{ij}$. Maximum likelihood estimation then
proceeds by maximization of
\[
\log \prod_{i=1}^{N}f \bigl( \boldsymbol{
\varepsilon }_{i} \bigl( \psi ^{d},\psi ^{s},\eta
,\alpha \bigr) |\mathbf{L}_{i},z;\omega \bigr) f (
\mathbf{Z}%
_{i}|\mathbf{L}_{i};\alpha )
\]%
with respect to $ ( \psi ^{d},\psi ^{s},\eta ,\alpha ,\omega  ) $.
A simple alternative to maximum likelihood estimation entails finding $%
\widehat{\alpha }$ that maximizes the partial log-likelihood
\[
\log \prod_{i=1}^{N}f (
\mathbf{Z}_{i}|\mathbf{L}_{i};\alpha )
\]%
and then finding the parameter value $ ( \widehat{\psi }^{d},\widehat{%
\psi }^{s},\widehat{\eta } ) $ that solves the following generalized
estimating equation with independence working correlation structure:
\[
\sum_{ij}\frac{\partial \varepsilon _{ij} ( \psi ^{d},\psi ^{s},\eta ,%
\widehat{\alpha } ) }{\partial  ( \psi ^{d},\psi ^{s},\eta  )
|_{\widehat{\psi }^{d},\widehat{\psi }^{s},\widehat{\eta }}}\varepsilon
_{ij} \bigl( \widehat{\psi }^{d},\widehat{\psi
}^{s},\widehat{\eta },\widehat{%
\alpha } \bigr) =0.
\]%
 Note that the dependence of $\varepsilon _{ij}$ on $ ( \delta
^{d},\delta ^{s} ) $ has been suppressed in the notation used above;
such dependence is made explicit in a sensitivity analysis which is obtained
by repeating either maximum likelihood estimation or the estimating
equations approach given above as $ ( \delta ^{d},\delta ^{s} ) $
is varied within a finite set of user-specified functions $\Gamma $ $=\{$ $%
\delta _{\lambda ^{d}}^{d},\delta _{\lambda ^{s}}^{s}\dvtx \lambda = (
\lambda ^{d},\lambda ^{s} ) \}$ indexed by a finite dimensional
parameter $\lambda $ with $ ( \delta _{0}^{d},\delta _{0}^{s} ) \in
$ $\Gamma $ corresponding to the ignorability assumption, that is, $\delta
_{0}^{d}=\delta _{0}^{s}\equiv 0$.

In applying this approach it is helpful to briefly describe possible
functional forms for the selection bias functions $\delta ^{d},\delta ^{s}$.
In practice, it may be convenient to specify simple parametric models for
each of these functions as illustrated in the following display. To
illustrate, one may set the function $g(\mathbf{Z}_{i(j)})=\sum_{j^{\prime
}\neq j}Z_{ij^{\prime }}$ to equal the number of exposed individuals in
cluster $i$ excluding person $j$:
\begin{eqnarray*}
\delta _{\lambda ^{d}}^{d,1} ( z,g,l,h ) &=&\lambda ^{d}z,
\quad\delta _{\lambda ^{s}}^{s,1} ( g,l,h ) =\lambda ^{s}g,%
\\
\delta _{\lambda ^{d}}^{d,2} ( z,g,l,h ) &=&z \bigl( \lambda
_{1}^{d}+\lambda _{2}^{d}g \bigr) ,
\\
\delta _{\lambda ^{s}}^{s,2} ( g,l,h ) &=&g \bigl( \lambda
_{1}^{s}%
+\lambda _{2}^{s}l
\bigr),
\end{eqnarray*}%
where for $\delta _{\lambda ^{d}}^{d,1}$ and $\delta _{\lambda ^{d}}^{d,2}$,
the scalar parameters $\lambda ^{d}$ and $ ( \lambda _{1}^{d},\lambda
_{2}^{d} ) $ encode the magnitude and direction of unmeasured
confounding for the effect of person $j$'s exposure $Z_{ij}$ on his outcome $%
Y_{ij}$, and for $\delta _{\lambda ^{s}}^{d,1}$ and $\delta _{\lambda
^{s}}^{d,2}$, the scalar parameters $\lambda ^{s}$ and $ ( \lambda
_{1}^{s},\lambda _{2}^{s} ) $ encode the magnitude and direction of
unmeasured confounding for the causal effect of the total number exposed $%
\sum_{j^{\prime }\neq j}Z_{ij^{\prime }}$ excluding person $j$, within the
cluster $i$ on person $j$'s outcome $ Y_{ij}$.

The functions $ \delta _{\lambda ^{d}}^{d,2}$ and $\delta _{\lambda
^{s}}^{s,2}$ model interactions between $Z_{ij}$ and $\sum_{j^{\prime }\neq
j}Z_{ij^{\prime }}$, thus allowing for heterogeneity in the selection bias
function. Since the functional form of $ ( \delta _{\lambda
^{d}}^{d},\delta _{\lambda ^{s}}^{s} ) $ is not identified from the
observed data, we generally recommend reporting results for a variety of
functional forms.
\end{appendix}

\section*{Acknowledgments}
Tyler J. VanderWeele was supported by NIH Grant R01
ES017876. Eric J. Tchetgen Tchetgen  was supported by NIH Grants R01ES020337, R01AI104459,
R21ES019712 and U54GM088558. M. Elizabeth Halloran was supported by NIH Grants R37AI032042
and R01AI085073.
The authors thank the editors and
reviewers for helpful comments.  The content is solely the responsibility of the authors
and does not necessarily reflect the official views of the National
Institute of Allergy and Infectious Diseases or the National Institutes of
Health.



\end{document}